\RequirePackage[OT1]{fontenc}
\documentclass[journal,10pt]{IEEEtran}
\IEEEoverridecommandlockouts

\usepackage[caption=false,font=footnotesize]{subfig}
\usepackage{verbatim}
\usepackage{cite}
\usepackage{array}
\usepackage{amsmath,amssymb,amsfonts}
\usepackage{amsthm}
\usepackage{bm}
\usepackage{algorithmic}
\usepackage{graphicx}
\usepackage{tikz}
\usepackage{circuitikz}
\usepackage{siunitx}
\usepackage{gensymb}
\usepackage{bbold}
\usepackage{soul}
\setstcolor{red}
\setulcolor{blue!80}
\setul{0.5ex}{1.5pt}
\usetikzlibrary{external}
\usetikzlibrary{shapes.geometric, arrows,chains, positioning}
\usetikzlibrary{arrows.meta} 
\usetikzlibrary { decorations.pathmorphing, decorations.pathreplacing, decorations.shapes, } 
\usetikzlibrary{spy}
\usepackage{pbox}
\usepackage{relsize}
\tikzset{
    pin/.style = {font = \relsize{-2}} % pin font size
}
\ctikzset{
    bipoles/length = 4em, % bipole size
    font = \relsize{-1}, % default font size
}
\usepackage{textcomp}
\usepackage{xcolor}
\def\BibTeX{{\rm B\kern-.05em{\sc i\kern-.025em b}\kern-.08em
    T\kern-.1667em\lower.7ex\hbox{E}\kern-.125emX}}
\usepackage{mathtools}
\usepackage{schemabloc}
\usetikzlibrary{circuits}
\usetikzlibrary{arrows}

\makeatletter
\let\MYcaption\@makecaption
\makeatother
\usepackage[font=footnotesize]{subcaption}
\makeatletter
\let\@makecaption\MYcaption
\makeatother
\usepackage[caption=false,font=footnotesize]{subfig}
\usepackage{caption}
\usepackage{pgfplotstable}
\usepackage{multirow}
\usepackage{mathtools}
\usepackage{commath}

\newcommand{\RNum}[1]{\uppercase\expandafter{\romannumeral #1\relax}}
\graphicspath{ {./confimages/} }
 \usepackage{pgfplots}
  \pgfplotsset{compat=newest}
  \usetikzlibrary{plotmarks}
  \usetikzlibrary{arrows.meta}
  \usepgfplotslibrary{patchplots}
  \usepackage{grffile}

\newtheorem{proposition}{Proposition}

\usepackage{fancyhdr,graphicx}
\usepackage[ruled,vlined]{algorithm2e}
\usepackage{hhline}
\usepackage{booktabs}
\usepackage{circuitikz}
\usepackage{siunitx}
\usepackage[inline]{enumitem}

\usepackage[hyphens]{url}
\usepackage{balance}
\usepackage[commandnameprefix=always]{changes}
\definechangesauthor[color=blue,name=Vassilis]{VP}
\definechangesauthor[color=cyan,name=Dimitris]{DK}

\usepackage{nicematrix}

\usepackage{hyperref}
\hypersetup{
    colorlinks=true,
    linkcolor=blue!70,
    filecolor=magenta,      
    urlcolor=cyan,
    citecolor=teal!77,
    pdftitle={RISvariationsJournal}
}

\usepackage{flafter} 

\makeatletter
\renewenvironment{proof}[1][\relax]{\par
  \pushQED{\qed}%
  \normalfont \topsep6\p@\@plus6\p@\relax
  \trivlist
  \item[\hskip\labelsep\itshape
    \ifx#1\relax \proofname\else\proofname{} of #1\fi\@addpunct{.}]\ignorespaces
}{%
  \popQED\endtrivlist\@endpefalse
}
\makeatother

\usepackage[breakable]{tcolorbox}
\newtcolorbox{abox}{
    boxrule = 1.5pt,
    colframe = blue!80,
    rounded corners,
    use color stack,
    breakable,
    arc = 5pt
}

\begin{document}

\allowdisplaybreaks % page breaks for equations

\title{RIS Beamforming under Element-Level Variations: Statistical Characterization and Robust Design}

\author{Dimitris Kompostiotis,~\IEEEmembership{Student~Member,~IEEE}, Dimitris Vordonis,~\IEEEmembership{Student~Member,~IEEE},\\ Vassilis Paliouras,~\IEEEmembership{Member,~IEEE}, and George C. Alexandropoulos,~\IEEEmembership{Senior~Member,~IEEE}
\thanks{D. Kompostiotis, D. Vordonis, and V. Paliouras are with the Electrical and Computer Engineering Department, University of Patras, 26504 Rio-Patras, Greece (e-mails: d.kompostiotis@ac.upatras.gr, d.vordonis@ac.upatras.gr, paliuras@upatras.gr).}
\thanks{G. C. Alexandropoulos is with the Department of Informatics and Telecommunications, National and Kapodistrian University of Athens, 16122 Athens, Greece and with the Department of Electrical and Computer Engineering, University of Illinois Chicago, IL 60601, USA (e-mail: alexandg@di.uoa.gr).}
\vspace{-0.33cm}}

\maketitle

\begin{abstract}
In this paper, a novel analytical framework to characterize the impact of element-level variations on the radiation characteristics of reconfigurable intelligent surfaces (RISs) is introduced. Specifically, a statistical model is proposed to capture the effects of varactor capacitance fluctuations on the RIS reflection coefficients, and, subsequently, on the resulting power radiation pattern; both low- and large-variance independent perturbation scenarios, are investigated. Leveraging the proposed statistical model, a low complexity greedy optimization methodology is presented, having the goal to optimize the expected RIS radiation power, thereby, generating inherently robust configurations. Furthermore, the analytical proposed model serves as an efficient alternative to computationally expensive Monte Carlo simulations, enabling the quantification of element sensitivity to manufacturing and operational tolerances. As demonstrated, optimizing the mean power pattern significantly enhances system performance under element-level variations. For typical RIS sizes (e.g., $32{\times}32$ or $64{\times}64$), a main lobe gain exceeding $2$~dB and a sidelobe suppression of approximately $10$~dB are achieved. %To the best of the authors' knowledge, this is the first work to explicitly incorporate element-wise capacitance fluctuations into the RIS power pattern model and to develop targeted optimization techniques mitigating their impact.    
\end{abstract}

\begin{IEEEkeywords}
Reconfigurable intelligent surfaces, reflection coefficient variations, stochastic modeling, power radiation pattern, RIS optimization.
\end{IEEEkeywords}

\section{Introduction and Motivation}
\label{sec:Intro_and_motivation}
\IEEEPARstart{T}{he} sixth generation (6G) of wireless networks is envisioned to enable a plethora of novel internet-of-things (IoT) applications, ranging from smart cities to autonomous vehicles~\cite{RIS_smart_cities,wymeersch2022localisation}. To support these advanced use cases, the concepts of enhanced mobile broadband (eMBB), ultra-reliable low latency communications (URLLC) and massive machine type communications (mMTC) are being reinforced~\cite{10005197, huang2019reconfigurable}, alongside critical capabilities, such as security, positioning, and sensing~\cite{10989512, kompostiotis2024evaluation, 9852716}. These advancements are facilitated by the integration of innovative hardware technologies into the physical layer, most notably massive multiple-input multiple-output (MIMO) systems and reconfigurable intelligent surfaces (RISs)~\cite{Tsinghua_RIS_Tutorial, bjornson2022reconfigurable, basar2023RIS_mag}. Following continuous progress in micro-electromechanical systems and metamaterials~\cite{basar2023RIS_mag}, specifically RISs have been developed as planar arrays capable of dynamically controlling the reflection of incident electromagnetic waves~\cite{9847080, alexandropoulos2023ris}. By transforming the propagation channel into a reconfigurable entity~\cite{strinati2021wireless, 9847080}, the wireless environment is effectively manipulated to meet specific network requirements. For instance, connectivity is improved by establishing virtual line-of-sight (LoS) paths to overcome blockages~\cite{RDK21}, and energy efficiency is maintained through the use of nearly passive reflecting elements that do not require power amplifiers~\cite{huang2019reconfigurable}. Also, physical-layer security is enhanced by optimizing reflections to mitigate eavesdropping and jamming~\cite{cui2019secure, guan2020intelligent, kompostiotis2023secrecy}, and localization and sensing accuracy is significantly boosted by addressing non-line-of-sight (NLoS) limitations~\cite{bjornson2022reconfigurable, RIS_loc,10243495}. 

The realization of the aforementioned applications and capabilities of RIS, relies heavily on its ability to perform accurate passive beamforming, which is inherently determined by the electromagnetic radiation pattern of the RIS. By optimizing the RIS configuration, the reflected energy is spatially shaped to meet specific service requirements. This precise control is fundamental for maximizing the received signal power in communication links, ensuring high angular resolution for sensing and localization tasks, and spatially filtering signals to enhance physical layer security~\cite{kompostiotis2025optimizing,rahal2022arbitrary}. To shape the RIS radiation pattern, analytical models are typically employed. Far-field models assume planar wavefronts, determining the pattern via directional angles~\cite{bjornson2022reconfigurable,ramezani2023broad, ramezani2023dual}, whereas near-field models account for spherical wavefront curvature by relying on the exact Euclidean distances from the elements to the target~\cite{liu2023near}. However, translating these theoretically computed configurations into practical ones~\cite{cai2020practical}, is hindered by hardware imperfections, such as manufacturing variations, phase-dependent amplitude responses, and mutual coupling, alongside noise and complex multipath effects~\cite{abeywickrama2020intelligent, kompostiotis2023received, bjornson2021optimizing, vordonis2025evaluating, kompostiotis2023secrecy}. Consequently, bridging the gap between ideal models and actual real-world behavior to precisely determine the required physical RIS configuration remains a significant challenge.

In many practical RIS implementations (e.g., varactor-based RISs~\cite{Rains_OpenRIS_Github}), the configuration of each unit cell is typically achieved by applying a bias voltage~\cite{11030582,Rains_OpenRIS_Github}, which alters the capacitance of the integrated varactor diode. However, the actual capacitance $C_i$ of each element is not strictly identical to its nominal target. Even when an identical DC bias voltage is applied across the entire RIS, the effective capacitance of the $i$th element could be modeled as $C_i^{\text{effective}} {=} C_i^{\text{nom}} {+} \Delta C_i$, where $C_i^{\text{nom}}$ is the expected nominal capacitance and $\Delta C_i$ represents random, element-specific deviations of the controllable capacitance.
These deviations ($\Delta C_i$) arise from two primary sources: \textit{i}) \textit{operational radio frequency (RF) voltage swings:} a significant factor that causes capacitance variation is the electromagnetic signal hitting the RIS itself. As analyzed in the literature, a varactor's junction capacitance is a nonlinear function of the applied voltage~\cite{buisman2012rf, akaike2004analysis}. The final capacitance does not only depend on the static DC bias voltage used to set the operating point~\cite{akaike2004analysis, buisman2012rf}, but it is also affected by the small dynamically changing alternating RF signal it receives~\cite{akaike2004analysis}. Since the electromagnetic wave does not illuminate the entire RIS uniformly (each element sees a different signal amplitude and phase), the resulting capacitance shift, naturally differs from element to element. Furthermore, research shows that conventional varactors can experience up to a 10\% capacitance variation solely due to these RF voltage swings under the exact same excitation~\cite{akaike2004analysis}. \textit{ii}) \textit{Manufacturing imperfections (process variations):} random capacitance deviations are inherently introduced by physical fabrication tolerances. Because these fluctuations are caused by numerous independent microscopic factors, their distribution is modeled as Gaussian, as dictated by the central limit theorem. This statistical behavior is directly corroborated by manufacturer datasheets. For instance, the Skyworks SMV1408 varactor~\cite{Skyworks_SMV1408_Datasheet} utilized in the OpenRIS architecture~\cite{Rains_OpenRIS_Github} presents an approximate 20\% capacitance deviation ($1.71$\,pF to $2.11$\,pF at $4$\,V), which is further exacerbated by parasitic packaging capacitance. Consequently, modeling the RIS element capacitance as a Gaussian random variable is both theoretically sound and practically necessary. 
Following the fabrication of a large number of devices, each may exhibit deviations from the nominal capacitance value, further exacerbated by the aforementioned external factors. In this context, to accurately predict the actual performance of an RIS, a statistical rather than a deterministic model is required. The proposed statistical framework evaluates the expected (mean) radiation pattern and its variance, offering a design tool that quantifies the performance degradation caused by manufacturing and environmental variations and facilitates component specification definition prior to large-scale deployment. Moreover, it enables the incorporation of the impact of random and time-varying parameter fluctuations during the design phase, reducing the need for potentially costly post-production compensation.

Therefore, existing variations in the physical parameters of the RIS elements~\cite{wang2025impact}, resulting in capacitance variations on varactor-based unit cells, translate into deviations of the intended reflection coefficients~\cite{wang2025impact}. These deviations manifest primarily as phase errors, and, to a lesser extent, as amplitude mismatches across the RIS aperture. Motivated by this, in this paper, we evaluate the extent to which the array factor deviates from the ideal configuration, triggering reduced beamforming gain, increased sidelobe levels, and potential beam pointing errors. Such impairments directly impact the aforementioned applications that rely on precise radiation control, including localization and sensing (where angular accuracy is degraded), codebook-based beam alignment~\cite{rahal2022arbitrary, kompostiotis2025optimizing}, where the optimal configuration may be misidentified, and high-throughput communication links, where reduced signal-to-noise ratio (SNR) and interference from sidelobes deteriorate the link quality. Therefore, such variations degrade the overall radiation pattern quality, highlighting the need for calibration or adaptive tuning mechanisms to mitigate their impact on system performance. 

In array processing literature~\cite{li2003robust,vorobyov2003robust,lorenz2005robust}, robust beamforming typically models macroscopic geometric uncertainties as variations in spatial steering vectors. While this approach can be conceptually generalized to RIS-aided systems, it is structurally distinct from the framework proposed herein. Although the hardware perturbations considered herein could be mapped into variations on the steering vector, the resulting variation distribution is non-Gaussian, in contrast with standard literature~\cite{li2003robust,vorobyov2003robust,lorenz2005robust}. To address this mismatch, variations are introduced directly at the physical hardware level, specifically into individual RIS element capacitances, where a Gaussian distribution can be naturally justified (as further elaborated in Section~\ref{sec:model_formulation}). By modeling how these capacitance shifts alter the complex reflection coefficients (i.e., the RIS configuration), a more realistic representation of hardware variations is achieved, enabling explicit compensation via statistical pattern modeling to ensure reliable practical deployments.
Therefore, this paper introduces an analytical framework for quantifying the impact of element-level hardware variations on the radiation characteristics of RISs, and its main contributions are summarized as follows:
\begin{itemize}
    \item A novel stochastic model is proposed, which captures the effect of controllable-varactor capacitance random perturbations to the RIS reflection coefficients and, consequently, on the resulting RIS power radiation pattern. 
    \item Leveraging the proposed model, an optimization scheme is adopted to optimize the expected RIS power pattern, yielding inherently robust RIS configurations, and explicitly compensating the element-level imperfections.
    \item The proposed approach quantifies system sensitivity to individual physical fluctuations of the RIS unit elements, enabling the derivation of tolerance bounds for component variations. Specifically, low- and large-variance capacitance perturbation scenarios are investigated per RIS element. By evaluating the variance of the RIS power pattern, the proposed model establishes a bound on the inherent pattern robustness that a specific RIS configuration can guarantee, thereby ensuring predictable link reliability during real-world deployment. 
    \item Finally, the presented numerical investigation showcases that the proposed statistical model for the RIS power radiation pattern, accounting for per-element capacitance variations, serves as an efficient alternative to Monte Carlo simulations, significantly reducing execution time. 
\end{itemize}
To the best of the authors' knowledge, this is the first work to explicitly incorporate element-wise capacitance fluctuations into the RIS power pattern model (serving also as an efficient Monte Carlo alternative), and to develop targeted optimization techniques mitigating their impact. 

The remainder of the paper is organized as follows. 
Section~\ref{sec:systemmodel} introduces the RIS reflection radiation pattern model, while Section~\ref{sec:model_formulation} justifies the Gaussian nature of RIS element capacitance variations and integrates these variations into the RIS power pattern, thereby establishing a robust stochastic model, that yields key statistical metrics, specifically the pattern's mean and variance. Section~\ref{sec:RIS_optimization} proposes an algorithm, to optimize the expected RIS power pattern while accounting for inherent element-level imperfections causing capacitance perturbations. Section~\ref{sec:simulation_results} provides the simulation results validating the proposed stochastic models and the proposed optimization technique, and finally Section~\ref{sec:conclusions} concludes the paper.

\textbf{Notation:} Bold lower case letters are used for column vectors (e.g., $\bm{x}$) and upper case ones for matrices (e.g., $\bm{X}$). To refer to the $(i,j)$th element, to the $i$th column, to the $i$th row of matrix $\bm{X}$ and to the $i$th element of vector $\bm{x}$  the notations $\bm{X}_{i,j}$, $\bm{X}_{:,i}$, $\bm{X}_{i,:}$ and $\bm{x}_i$ are used respectively. The transpose, conjugate transpose, and the conjugate operators are denoted by $(\cdot)^\mathsf{T}$, $(\cdot)^\mathsf{H}$ and $(\cdot)^\mathsf{*}$ respectively. For a complex-valued scalar $x$, $|x|$ and $\angle x$ denote its absolute value and phase, respectively. 
$\mathbb{C}^{M {\times} N}$ and $\mathbb{R}^{M {\times} N}$
denote the set of $M {\times} N$ complex-valued and real-valued matrices respectively, while $\odot$ implies the Hadamard product. Moreover, $\lfloor{x\rfloor}$ denotes the integer part of $x$ and the operators $\Re$ and $\Im$ denote the real and imaginary part of a complex math object (e.g., matrix, vector, scalar) respectively. The notation $\{\cdot\}$ is used to denote a set of discrete elements, explicitly listing all its members within braces. Conversely, the notation $[\cdot,\cdot]$ or $(\cdot,\cdot)$ represents a continuous interval, closed or open at its boundaries, including all values between the specified endpoints. Notation $\operatorname{erf}(\cdot)$ and $\operatorname{erfc}(\cdot)$ denote the error function and the complementary error function, respectively. Finally, $\mathbb{E}_{\bm{x}}$ denotes the expected value operator, with $\bm{x}$ representing the variables with respect to which the integration is performed, to compute the mean value and $\jmath{=}\sqrt{{-}1}$ denotes the imaginary unit. 

\section{Considered RIS Response Modeling}
\label{sec:systemmodel}
In this paper, an RIS, which consists of $N {=} N_V{\times}N_H$ elements is considered. These elements are deployed on a two-dimensional rectangular grid with $N_V$ elements per column and $N_H$ elements per row with an inter-element spacing distance equal to $d_V$ for the vertical and $d_H$ for the horizontal component, respectively. The area of each element is therefore given by $d_Vd_H$. The specific values of $d_V$ and $d_H$ have been chosen based on the RIS's center operating frequency, denoted by $f_c$. Without loss of generality, the RIS is assumed to be positioned at the $yz$-plane (i.e., $x{=}0$), and its elements are indexed row-by-row by $n{=}1,2,{\ldots},N$. Thus, the location of the $n$th element is given by
\begin{equation}
    \bm{U}_{:,n} = \big[0, \hspace{0.1cm} \mathrm{i}(n)d_H, \hspace{0.1cm} \mathrm{j}(n)d_V\big]^{\mathsf{T}},
    \label{e:ris_location}
\end{equation}
where $\mathrm{i}(n) {=} \operatorname{mod}(n{-}1, N_H)$ and $\mathrm{j}(n) {=} \lfloor(n{-}1) / N_H\rfloor$ are the horizontal and vertical indices of element $n$, on the two-dimensional grid, respectively~\cite{bjornson2021optimizing}. As a result, the matrix $\bm{U}{\in}\mathbb{R}^{3 {\times} N}$ represents the positions of all RIS elements relative to the selected reference coordinate system.

The signal propagation medium towards the RIS is commonly modeled via the transmission line model (TLM)~\cite{abeywickrama2020intelligent}, characterized by the free-space impedance $Z_0{=}377~\Omega$~\cite{costa2021electromagnetic,kompostiotis2023secrecy,abeywickrama2020intelligent}. Upon impingement of the electromagnetic (EM) wave on an RIS element, the incident signal is partitioned into reflected, refracted, and scattered components. The power distribution among these components is determined by~\cite{costa2021electromagnetic}: \begin{enumerate*}[label=(\alph*)] \item the incident signal wavelength, \item the incidence angle, \item the surface material properties, and \item the surface geometry \end{enumerate*}. However, since RISs are designed to minimize scattering and absorption~\cite{costa2021electromagnetic,wu2021intelligent}, and refracted waves are irrelevant for receivers located in the reflection half-space  (i.e., in front of the RIS), only the reflected power is primarily evaluated. 

To compute the total reflected field, Maxwell's equations are solved under boundary conditions dictated by the dynamically changing RIS configuration (i.e., the vector
containing each RIS-element’s reflection state). Consequently, the received signal is a superposition of individual element reflections~\cite{9847080,pan2020multicell}. Each element's contribution is quantified by its complex reflection coefficient, which dictates both the reflected amplitude and phase shift. By adopting a circuit model~\cite{costa2021electromagnetic,kompostiotis2023received}, each element is treated as an RLC load terminating the transmission line, with a complex impedance given by
\begin{equation}
Z(C_n, f) = \frac{\jmath 2 \pi f L_1 \left ( \jmath 2 \pi f L_2 + (\jmath 2\pi f C_n)^{{-}1} + R_1   \right) }{\jmath 2\pi f L_1 + \jmath 2 \pi f L_2 + (\jmath 2\pi f C_n)^{{-}1} + R_1},
\label{eq:load_impedance}
\end{equation}
where  $L_1$, $L_2$ and $R_1$, depend on the RIS construction technology~\cite{bjornson2021optimizing,costa2021electromagnetic} and $f$ denotes the frequency of interest.
Each reflecting element is programmed via an access point (AP)-controlled RIS controller that dynamically adjusts each varactor's capacitance $C_n$. Although 1-bit RIS operation is common in practice~\cite{alexandropoulos2023ris,RIS_1bit,bjornson2021optimizing,MIMO_1bit}, where $\bm{c} {=} [C_1,C_2,\ldots,C_N]^{\mathsf{T}}$ and $C_n {\in} \{C_{\text{on}},C_{\text{off}}\}, {\forall} n{=}1,2,{\ldots},N$, this work outlines a generalized analysis for an arbitrary number of resolution bits. Finally, each RIS element's reflection coefficient is given by
\begin{equation}
\omega_n(C_n) {=} \frac{Z(C_n, f) {-} Z_0}{Z(C_n, f) {+} Z_0} {=} \alpha(C_n) e^{\jmath\theta(C_{n})}, n{=}1,2,{\ldots},N,
\label{eq:refl_coeff}
\end{equation}
where $Z(C_n, f)$ is given by~\eqref{eq:load_impedance}, $Z_0$ is the free-space impedance and $\alpha(C_n){\in}[0,1]\forall n$, $\theta(C_n){\in}[0,\pi]\forall n$ are the $n$th RIS element's amplitude and phase response around a frequency $f$, respectively. Thus, each RIS element reflects the incident wave with an individual amplitude and phase shift, enabling constructive or destructive confluence to shape the spatial power distribution. For an RIS configuration vector $\bm{\omega}$, the RIS power radiation pattern is given by~\cite{ramezani2023broad}, as follows
\begin{equation}
    \text{A}(\varphi, \vartheta) = |\bm{\omega}^{\mathsf{T}}\,(\mathbf{a}_{\text{RIS}}(\varphi_{\text{AoA}},\vartheta_{\text{AoA}}) \odot \mathbf{a}^{*}_{\text{RIS}}(\varphi, \vartheta))|^2,
    \label{eq:power_factor}
\end{equation}
where $(\varphi, \vartheta)$ refers to a specific azimuth-elevation angle pair where the RIS reflects power, $(\varphi_{\text{AoA}},\vartheta_{\text{AoA}})$ is the AoA angle pair from which the transmitted signal approaches the RIS and \scalebox{0.92}{$\bm{\omega} {=} \Big[\alpha(C_1)e^{\jmath\theta(C_{1})}, \ldots , \alpha(C_N)e^{\jmath\theta(C_{N})}\Big]^{\mathclap{\mathsf{T}}} \in\mathbb{C}^N$}, is the RIS configuration. The $\mathbf{a}_{\text{RIS}}(\varphi,\vartheta)$ is the array response vector for both reception and transmission of the RIS~\cite{bjornson2022reconfigurable,kompostiotis2024evaluation} and specifically, $\mathbf{a}_{\text{RIS}}(\varphi_{\text{AoA}},\vartheta_{\text{AoA}})$ and $\mathbf{a}_{\text{RIS}}(\varphi,\vartheta)$ in~\eqref{eq:power_factor}, refer to the RIS-reception and the RIS-transmission array response vector, respectively. For the far-field case, it is given by
\begin{equation}
\scalebox{1}{$
\begin{aligned}
    \mathbf{a}_{\text{RIS}}(\varphi, \vartheta) = \Big[e^{-\,\jmath\mathbf{k}(\varphi, \vartheta)^\mathsf{T}\bm{U}_{:,1}}, \ldots, e^{-\,\jmath\mathbf{k}(\varphi, \vartheta)^\mathsf{T}\bm{U}_{:,N}}\Big]^{\mathsf{T}}.
\end{aligned}$}
    \label{e:array_response_vec}
\end{equation}
In~\eqref{e:array_response_vec}, $\bm{U}_{:,i}$ is the RIS's $i$th transceiving element position and the wave vector $\mathbf{k}(\varphi, \vartheta) {\in} \mathbb{R}^{3{\times} 1}$ is given by
\begin{equation}
\scalebox{0.97}{$
\begin{aligned}
    \mathbf{k}(\varphi, \vartheta) = \frac{2\pi}{\lambda} \Big[\cos{(\vartheta)}\cos{(\varphi)}, \cos{(\vartheta)}\sin{(\varphi)}, \sin{(\vartheta)}\Big]^{\mathsf{T}},
\end{aligned}$}
    \label{e:wave_vector}
\end{equation}
where $\lambda$ is the impinging on the RIS plane-wave's wavelength. The RIS array response vector in~\eqref{e:array_response_vec} accounts for both surface geometry and per-element radiation patterns. Although an isotropic pattern is assumed here for simplicity, the performance evaluation presented later on in Section~\ref{sec:simulation_results} incorporates the realistic directional patterns from~\cite{ramezani2023broad,ramezani2023dual}. 

The deterministic model established in Section~\ref{sec:systemmodel} characterizes the RIS radiation pattern in~\eqref{eq:power_factor} under ideal operating conditions. However, practical RIS deployments are inherently susceptible to hardware variations that deviate from this nominal baseline. To bridge this gap and evaluate realistic performance, the deterministic formulation must be extended into a stochastic model. Due to practical impairments like manufacturing tolerances and random noise~\cite{buisman2012rf,akaike2004analysis,Skyworks_SMV1408_Datasheet} (presented in Section~\ref{sec:Intro_and_motivation}) the RIS power pattern becomes a random function of the perturbed capacitances $\bm{c}+\bm{\Delta c}$ and is denoted as $\tilde{A}(\varphi, \vartheta)$. To characterize the resulting beamforming degradation, due to capacitance perturbations, Section~\ref{sec:model_formulation} derives the expected radiation pattern $\mathbb{E}_{\bm{\Delta c}}[\tilde{A}(\varphi, \vartheta)]$ and its standard deviation under Gaussian variations $\bm{\Delta c} {\in} \mathbb{R}^N$. Crucially, these statistical metrics provide the mathematical foundation to optimize the RIS configuration, thereby directly mitigating the variation-induced performance degradation in RIS power patterns.

\section{RIS Radiation Pattern Statistical Modeling}
\label{sec:model_formulation}
As aforementioned, the resulting RIS power pattern $\tilde{A}(\varphi, \vartheta)$ becomes a random
function of the perturbed capacities $\mathbf{c}{+}\bm{\Delta}\mathbf{c}$~\cite{Rains_OpenRIS_Github,Skyworks_SMV1408_Datasheet}, where $\bm{\Delta}\mathbf{c} {\in} \mathbb{R}^{N}$ denotes the perturbation term and $\bm{c}{\in} \mathbb{R}^{N}$ the initially configured (nominal) capacities. Since this can negatively impact the RIS beamforming capabilities and in order to quantify the effect of random varactor capacitance variations on the RIS power pattern, this section derives the expected radiation pattern and its variance for a Gaussian distributed perturbation $\bm{\Delta}\mathbf{c}$. The adoption of the mean-variance based framework is justified by two primary factors. First, although hardware imperfections become static post-fabrication, their exact per-element values are unknown a priori and practically unobservable due to prohibitive calibration overhead for massively produced RIS instances. And secondly, real-time stochasticity from bias circuit noise and thermal drifts makes a statistical analysis imperative.

\subsection{Proposed Additive Gaussian Capacitance Variations}
\label{appen_sec:RIS_Pattern_Incl_Var}
Starting from~\eqref{eq:power_factor}, $\text{A}(\varphi, \vartheta)$ is re-written as 
\begin{align}
    \text{A}(&\varphi, \vartheta) = |\bm{\omega}^{\mathsf{T}}\,(\mathbf{a}_{\text{RIS}}(\varphi_{\text{AoA}},\vartheta_{\text{AoA}}) \odot \mathbf{a}^{*}_{\text{RIS}}(\varphi, \vartheta))|^2 \\
    %&= |\bm{\omega}^{\mathsf{T}}\,\bm{r}(\varphi, \vartheta,\varphi_{\text{AoA}},\vartheta_{\text{AoA}})|^2 \\
    &= \bm{\omega}^{\mathsf{T}}\bm{r}(\varphi, \vartheta,\varphi_{\text{AoA}},\vartheta_{\text{AoA}}) \bm{r}^{\mathsf{H}}(\varphi, \vartheta,\varphi_{\text{AoA}},\vartheta_{\text{AoA}})\bm{\omega}^{*}\, \label{eq:init_matrix_form_pattern} \\
    & = \bm{\omega}^{\mathsf{T}}\,\bm{R}(\varphi, \vartheta,\varphi_{\text{AoA}},\vartheta_{\text{AoA}})\,\bm{\omega}^{*}\,,
    %\label{eq:power_factor}
\end{align}
where $\bm{r}(\varphi, \vartheta,\varphi_{\text{AoA}},\vartheta_{\text{AoA}}){=}\mathbf{a}_{\text{RIS}}(\varphi_{\text{AoA}},\vartheta_{\text{AoA}}) \odot \mathbf{a}^{*}_{\text{RIS}}(\varphi, \vartheta)$ and $\bm{R}(\varphi, \vartheta,\varphi_{\text{AoA}},\vartheta_{\text{AoA}}){=}\bm{r}\bm{r}^\mathsf{H} {\in} \mathbb{C}^{N{\times}N}$. Applying~\eqref{eq:refl_coeff} for the $\bm{\omega}_n$, $n{=}1,\ldots,N$ to~\eqref{eq:init_matrix_form_pattern}, and performing some algebraic manipulations, the RIS power pattern, with respect to (w.r.t.) the per RIS-element capacitance $C_i$, $i{=}1,\ldots,N$, is re-expressed as
\begin{align}
% \scalebox{0.94}{$
% \begin{aligned}
A(\varphi,\vartheta) &{=} 
% &\smash[b]{\sum_{i=1}^N} \! \alpha^2(C_i) |r_i(\varphi,\vartheta)|^2 
%        {+} \notag\\
%        &\quad 2 \sum_{i=1}^{N-1} \sum_{j{=}i{+}1}^N \!\!\!\! \Big(\alpha(C_i)\alpha(C_j) |r_i(\varphi,\vartheta)|
%           |r_j(\varphi,\vartheta)| \notag \\ 
%           &\quad\quad \cos\big(\theta(C_i) {-} \theta(C_j) + \angle r_i(\varphi,\vartheta) {-} \angle r_j(\varphi,\vartheta) \big) \kern-2pt\Big)\notag\\= 
           \sum_{i=1}^{N} \alpha^2(C_i) + 2 \sum_{i=1}^{N-1} \sum_{j{=}i+1}^N \Big( \alpha(C_i)\alpha(C_j) \notag \\
         &  {\cdot} \cos\kern-1pt\big(\theta(C_i) {-} \theta(C_j) {+} \angle r_i(\varphi,\vartheta) {-} \angle r_j(\varphi,\vartheta) \big) \kern-2pt\Big)\mathclap{,}
%\end{aligned}$}
\label{eq:rad_pattern}
\end{align}
where~\eqref{eq:rad_pattern} holds since $|r_i(\varphi,\vartheta)|{=}1, \forall i$. Consequently, the RIS power pattern, $\tilde{A}(\varphi,\vartheta)$, which considers the capacitance variation $\Delta C_i$ at the $i$th RIS element, is given by
\begin{align}
% \scalebox{0.95}{$
% \begin{aligned}
\tilde{A}(\varphi,\vartheta)&{=} \!\!\sum_{i=1}^N \! \alpha^2(C_i {+} \Delta C_i) {+} 2\! \sum_{i=1}^{N-1} \! \sum_{j{=}i{+}1}^{N} \!\!\! \Big(\! \alpha(C_i {+} \Delta C_i) \notag \\ &\kern10pt{\cdot}\,\alpha(C_j {+} \Delta C_j) 
\cos\!\big(\theta(C_i{+}\Delta C_i) {-} \theta(C_j{+}\Delta C_j) \notag\\
&\qquad\qquad\qquad +
\angle r_i(\varphi,\vartheta) {-} \angle r_j(\varphi,\vartheta)\big)\kern-2pt \Big),
%\end{aligned}$}
\label{eq:tppp}
\end{align}
where the perturbations $\Delta C_1, \Delta C_2, \ldots, \Delta C_N$ are assumed to be independent identically distributed (i.i.d.) Gaussian random variables with $\Delta C_i {\sim} \mathcal{N}(0,\sigma_c^2), \forall i$, representing the variation of each RIS-element's controllable capacitance.

Although a detailed device-level physical analysis is outside the scope of this paper, a brief evaluation is conducted to justify the Gaussian nature of the RIS elements' capacitance variations. To do so, the varactor is represented by the lumped-element model from~\cite{abeywickrama2020intelligent}, which accounts for packaging and interconnection parasitics~\cite{Skyworks_SMV1408_Datasheet,abeywickrama2020intelligent} using here $L_1 {=} 2.8\,\text{nH}$, $L_2 {=} 0.8\,\text{nH}$, and $R_1 {=} 1\,\Omega$~\cite{kompostiotis2023received}. Thus, manufacturing variations are introduced into the core diode element, whose voltage-dependent junction capacitance is expressed as
\begin{equation}
C_i(V) = C_{i0}\big(1 - {V}\,{V_i}^{-1}\big)^{-M_{gc}}.
\label{eq:diode}
\end{equation}
Among the parameters $C_{i0}$ (zero-bias junction capacitance), $V_i$ (junction potential), $V$ (applied DC tuning voltage) and $M_{gc}$ (grading coefficient) of~\eqref{eq:diode}, $C_{i0}$ dominates variation due to its high susceptibility to photolithographic area tolerances. A Gaussian distribution is assumed for all parameters, justified by the central limit theorem for aggregate fabrication imperfections and aligned with industrial foundry standards like the SkyWater PDK~\cite{skywater_pdk}. Consequently, an LTspice Monte Carlo simulation of $10000$ runs is executed. To reflect realistic hardware variations, a dominant variance of $\sigma {=} 10\%$ is assigned to $C_{i0}$, while tighter bounds are applied to $V_i$ ($\sigma {=} 2\%$) and $M_{gc}$ ($\sigma {=} 1\%$) due to logarithmic doping dependencies and precise epitaxial profiles, respectively. Also, a minimal $\sigma {=} 0.5\%$ is allocated to the regulated tuning voltage $V$, reflecting a highly regulated DC bias network. After performing the simulation, a Gaussian like distribution is obtained for both the $C_{\text{off}}$ ($0.48\,\text{pF}$ at $V {=} 2.5\,\text{V}$) and $C_{\text{on}}$ ($0.63\,\text{pF}$ at $V {=} 10\,\text{V}$) states. This outcome is fundamentally justified by the fact that $C_{i0}$ acts as the dominant variation factor (as aforementioned), coupled with the strict linear dependence of the total capacitance $C_i(V)$ on $C_{i0}$ as dictated by~\eqref{eq:diode}. Thus, this analysis establishes the mathematical foundation for both the Gaussian nature of RIS elements' capacitance variations and the subsequent phase-error and yield analysis.

\subsection{Expected RIS Power Radiation Pattern Characterization}
To simplify the analysis and facilitate the calculation of the expected value and the variance of the RIS power pattern, the RIS reflection coefficient is modeled as a (piecewise) linear function of capacitance. Consequently, the TLM is linearized. This linearization is primarily essential for the phase response, since the amplitude remains largely insensitive to capacitance variations (as illustrated in Fig.~\ref{fig:piecewise_lin_approx_TLM_tikz}) and can be approximated by its variation-free nominal value. Thus, the radiation pattern in~\eqref{eq:tppp} is rewritten as
\begin{align}
% \scalebox{0.95}{$
% \begin{aligned}
\tilde{A}(\varphi,\vartheta)& \approx \!\!\sum_{i=1}^N \! \alpha^2(C_i) {+} 2\! \sum_{i=1}^{N-1} \! \sum_{j{=}i{+}1}^{N} \!\!\! \Big(\! \alpha(C_i) \alpha(C_j) \notag \\&  \kern43pt
\cdot \cos\!\big(\theta(C_i{+}\Delta C_i) {-} \theta(C_j{+}\Delta C_j) \notag\\
&\qquad\qquad\qquad +
\angle r_i(\varphi,\vartheta) {-} \angle r_j(\varphi,\vartheta)\big)\kern-2pt \Big).
%\end{aligned}$}
\label{eq:RIS_pattern_with_variations}
\end{align} 
The TLM response in~\eqref{eq:refl_coeff} is initially linearized around the nominal region (Fig.~\ref{fig:piecewise_lin_approx_TLM_tikz} green plot) including both $C_{\text{on}}$ and $C_{\text{off}}$. Dictated by the limits of validity of the linear approximation, this model (adopted in Section~\ref{subsec:Exp_value_iid}), is restricted to small capacitance fluctuations, and specifically for a standard deviation up to approximately $5\%$ of the nominal value, $C_{\text{off}}$. Beyond this threshold, the linear model cannot capture the actual TLM behavior, necessitating the piecewise linear approximation shown in Fig.~\ref{fig:piecewise_lin_approx_TLM_tikz} (red plot) and adopted in Section~\ref{subsec:Exp_value_high_variations}.

\subsubsection{Expected Value of the RIS Power Radiation Pattern, under
small-variance Variations}\label{subsec:Exp_value_iid}
\begin{figure}[t]
    \centering
    \includegraphics{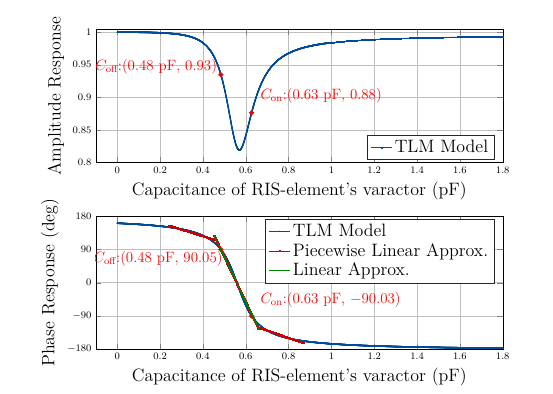}
    \caption{Linear and piecewise linear approximations of TLM-based RIS-element's phase response w.r.t. capacitance.}
    \label{fig:piecewise_lin_approx_TLM_tikz}
\end{figure}
In small capacitance variations scenarios, the phase reflection responses of the RIS elements are approximated using linear functions, that are given by
\begin{align}
    \theta(C_n) &= a_2 C_n + b_2,
\end{align}
where $a_2, b_2$ are parameters computed to fit a straight line to the nonlinear reflection model of each RIS element. Thus, the nonlinear model, governing the amplitude and phase of the reflection coefficient w.r.t. capacitance, is locally linearized in the nominal region using the least-squares (LS) method via the \texttt{polyfit} function of MATLAB\textsuperscript{\texttrademark}. This LS optimization must be re-executed if the circuit parameters ($L_1, L_2, R_1$) are adjusted to match a specific physical RIS element. For an optimized practical implementation with $L_1 {=} 2.8$~nH, $L_2 {=} 0.8$~nH, and $R_1 {=} 1\,\Omega$ from~\cite{bjornson2021optimizing,kompostiotis2023received} (Fig.~\ref{fig:piecewise_lin_approx_TLM_tikz}, green plot), the resulting linear phase coefficients are $a_2 {=} {-}1.22 {\times} 10^{15}$ and $b_2 {=} 678.33$. Consequently, the expected value of the RIS power pattern with respect to the perturbations $\bm{\Delta \bm{c}}$ is given by
\begin{align}
\mathbb{E}_{\bm{\Delta c}}&[\tilde{A}(\varphi,\vartheta)] {=}
\sum_{i=1}^N \mathbb{E}_{\Delta C_i}\big[\alpha^2(C_i)\big] \,\,  \notag\\  
& \!\! + 2 \smash[t]{\sum_{i=1}^{N-1}}\smash[t]{\sum_{j{=}i{+}1}^N} \mathbb{E}_{\Delta C_i, \Delta C_j}\Big[
\alpha(C_i) \alpha(C_j) \notag\\ &\quad
 {\cdot} \cos\!\big(\theta(C_i{+}\Delta C_i){-}\theta(C_j{+}\Delta C_j) {+} q_{i,j}(\varphi,\vartheta)\big)\Big],
\label{eq:mean_value_pattern_inti}
\end{align}
where $q_{i,j}(\varphi, \vartheta){=} \angle r_i(\varphi,\vartheta) {-} \angle r_j(\varphi,\vartheta)$ is constant for a specific AoA and AoD of the signal impinging on and reflected from the RIS, respectively. Also, the integration for the mean value computation in~\eqref{eq:mean_value_pattern_inti} is performed w.r.t. $\bm{\Delta c}{=}[\Delta C_1, \ldots, \Delta C_N]^T$ random variables, whose probability density function (PDF) is given by
\begin{equation}
f(\Delta C_i) = \frac{1}{\sqrt{2\pi\sigma_c^2}} \exp\!\left(-\frac{\Delta C_i^2}{2\sigma_c^2}\right), \,\, \forall \,i{=}1,\ldots, N.
\label{eq:gaussian_pdf_capac}
\end{equation}
Since the first term of the summation in~\eqref{eq:mean_value_pattern_inti} is independent of capacitance variations, it yields $\mathbb{E}_{\Delta C_i}\!\big[\alpha^2(C_i)\big] {=} \alpha^2(C_i)$.

Next, considering the cross-variable term (second term) in~\eqref{eq:mean_value_pattern_inti} and assuming that $\Delta C_1, \Delta C_2, \ldots, \Delta C_N$ are i.i.d. random variables with small variance (thus the linear approximation for RIS-element's reflection coefficient is adopted), the expectations of the second term are expressed as 
\begin{align}
&\mathbb{E}_{\Delta C_i, \Delta C_j}\!\big[\alpha(C_i)\alpha(C_j)\cos\!\big(\theta(C_i{+}\Delta C_i)  \notag \\ & \kern93pt -\theta(C_j{+}\Delta C_j) {+} q_{i,j}(\varphi,\vartheta)\big)\big] \notag\\
& {=} \alpha(C_i)\alpha(C_j) \mathbb{E}_{\Delta C_i, \Delta C_j}\big[
\cos\!\big( a_2\Delta C_i {-} a_2\Delta C_j {+} q^{\prime}_{i,j} \big)\big],
\label{eq:cross_term_mean_value}
\end{align}
where $q^{\prime}_{i,j}{=} q_{i,j} {+} a_2C_i {+} b_2 {-}
a_2C_j {-} b_2, \! {\forall}(i,j)$ are constants w.r.t. $\bm{\Delta c}{=}[\Delta C_1, \ldots, \Delta C_N]^T$; however they still depend on the nominal (i.e., without variation) capacitance of each RIS element. Expectations in~{\eqref{eq:cross_term_mean_value}} are further expanded as
\begin{align}
%\scalebox{0.9}{$\begin{aligned}
   &\alpha(C_i)\alpha(C_j) \mathbb{E}_{\Delta C_i, \Delta C_j}\big[
\cos\!\big( a_2\Delta C_i {-} a_2\Delta C_j {+} q^{\prime}_{i,j} \big)\big] \notag \\
& {=} \alpha(C_i)\alpha(C_j) \mathbb{E}_{\Delta C_i, \Delta C_j}\Big[ \Re \Big\{
e^{\jmath \big( a_2\Delta C_i {-} a_2\Delta C_j {+} q^{\prime}_{i,j} \big)}\Big\}\Big] \notag \\& {=} \alpha(C_i)\alpha(C_j) \Re \Big\{ \mathbb{E}_{\Delta C_i, \Delta C_j}\Big[
e^{\jmath \big( a_2\Delta C_i {-} a_2\Delta C_j {+} q^{\prime}_{i,j} \big)}\Big] \Big\} \notag \\ & {=} \alpha(C_i)\alpha(C_j) \Re \! \Big \{ e^{\jmath  q^{\prime}_{i,j}} \mathbb{E}_{\Delta C_i}\!\Big[ e^{\jmath  a_2\Delta C_i} \Big]\!\mathbb{E}_{\Delta C_j}\!\Big[ e^{{-}\jmath a_2 \Delta C_j} \Big] \!\Big \} \label{eq:2nd_term_intermediate_staep}\\&{=}\alpha(C_i)\alpha(C_j) \cos (q^{\prime}_{i,j})e^{-a^{2}_2 \sigma^{2}_{c}} \mathclap{,}
%\end{aligned}$}
\label{eq:2nd_term_mean_value}
\end{align}
where~\eqref{eq:2nd_term_intermediate_staep} holds since $\Delta C_i$ and $\Delta C_j$ are i.i.d. random variables, while~\eqref{eq:2nd_term_mean_value} follows by computing the corresponding mean values, as one-dimensional integrals due to expected value definition. This computation leads to $\mathbb{E}_{\Delta C_i}\!\big[ e^{\jmath  a_2\Delta C_i} \big]{=}\mathbb{E}_{\Delta C_j}\!\big[ e^{{-}\jmath a_2 \Delta C_j} \big]{=}e^{-\frac{1}{2}a_2^2\sigma_c^2}$ as shown in Appendix~\ref{appendix:1}. Finally, $\mathbb{E}_{\bm{\Delta c}}[\tilde{A}(\varphi,\vartheta)]$ is obtained by substituting%~\eqref{eq:1st_term_mean_value} and
~\eqref{eq:2nd_term_mean_value} into~\eqref{eq:mean_value_pattern_inti} and is given by \textbf{Proposition~\ref{proposition:iid_gaussian_case}}.
\begin{proposition}\label{proposition:iid_gaussian_case}
The expectation of the RIS power radiation pattern, {$\tilde{A}(\varphi,\vartheta)$}, in the presence of small-variance i.i.d. Gaussian capacitance variations, is given by
\begin{align}
% \scalebox{1}{$
% \begin{aligned}
        \mathbb{E}_{\bm{\Delta c}}&[\tilde{A}(\varphi,\vartheta)] {=}  \sum_{i{=}1}^{N} \! \alpha^2(C_i) \notag \\ & \kern17pt + 2 \sum_{i=1}^{N-1} \! \sum_{j{=}i{+}1}^N \!\! \Big( \alpha(C_i)\alpha(C_j) \cos (q^{\prime}_{i,j}) e^{-\alpha_2^2\sigma_c^2}\Big).
%\end{aligned}$}
\label{eq:mean_pattern_final}
\end{align}
\begin{proof}
    The result follows directly from Section~\ref{subsec:Exp_value_iid}.
\end{proof}
\end{proposition}
Proposition~\mbox{\ref{proposition:iid_gaussian_case}} shows that the mean RIS power radiation pattern depends on the nominal capacitance per RIS element and on some constant parameters. The parameters determining the RIS power radiation pattern are associated with the linearization parameters of the phase of the reflection coefficient, as well as with the variance of the distribution describing the per-element capacitance fluctuations. 

\subsubsection{Expected Value of the RIS Power Radiation Pattern, under large-variance Variations} \label{subsec:Exp_value_high_variations} 
In scenarios characterized by larger capacitance variations, the effective capacitance ($C_i{+}\Delta C_i, \, \forall i$) of each RIS element falls outside the linearized operation region, initially assumed in Section~\ref{subsec:Exp_value_iid}. Consequently, the linear approximation of the reflection coefficient's phase, becomes inaccurate. To address this limitation, the reflection coefficient's phase response is approximated using a piecewise linear function
\begin{align}
    \theta(C_n) &= \begin{cases} 
        a_{21} C_n + b_{21}, & \text{if } C_n < c_1 \\
        a_{22} C_n + b_{22}, & \text{if } c_1 \leq C_n \leq c_2, \\
        a_{23} C_n + b_{23}, & \text{if } C_n > c_2
    \end{cases}
    \label{eq:piecewise_func}
\end{align}
where $a_{21}$,  $b_{21}$, $a_{22}$,  $b_{22}$, $a_{23}$,  $b_{23}$ denote the coefficients of the piecewise linear model. The phase of the reflection model is approximated across its operating range using a piecewise linear function derived via the LS method using again the \texttt{polyfit} function of MATLAB\textsuperscript{\texttrademark}. Again, if the circuit parameters ($L_1$, $L_2$ and $R_1$) are adjusted to match a specific physical RIS element, the LS-based piecewise linearization process should be re-executed for each linearization interval. For an implementation with $L_1 {=} 2.8\text{~nH}$, $L_2 {=} 0.8\text{~nH}$, and $R_1 {=} 1\,\Omega$~\cite{bjornson2021optimizing,kompostiotis2023received} (Fig.~\ref{fig:piecewise_lin_approx_TLM_tikz}), the intervals in~\eqref{eq:piecewise_func} are defined by $c_1 {=} 0.45\text{~pF}$ and $c_2 {=} 0.65\text{~pF}$ and the resulting phase coefficients are $a_{21} {=} {-}1.81 {\times} 10^{14}$, $b_{21} {=} 198.76$, $a_{22} {=} {-}1.22 {\times} 10^{15}$, $b_{22} {=} 678.33$, $a_{23} {=} {-}1.20 {\times} 10^{14}$, and $b_{23} {=} 10.61$. The amplitude response is assumed independent of capacitance variations due to its negligible deviation from the nominal value.

Under the assumption of piecewise linear phase response $\theta(C_n{+}\Delta C_n)$ in~\eqref{eq:piecewise_func}, the expected RIS power radiation pattern is still determined by~\eqref{eq:mean_value_pattern_inti}. The expectation integrates over the i.i.d. random vector $\bm{\Delta c} {=} [\Delta C_1, \ldots, \Delta C_N]^T$, which follows the Gaussian PDF of~\eqref{eq:gaussian_pdf_capac} with an increased variance $\sigma_c^2$. The first summation term remains $\mathbb{E}_{\Delta C_i}\big[\alpha^2(C_i)\big] {=} \alpha^2(C_i)$, while, the expectation of the cross-term in~\eqref{eq:mean_value_pattern_inti}, is given by 
\begin{align}
    \phantom{x} &\kern-10pt \mathbb{E}_{\Delta C_i, \Delta C_j}\Big[
\alpha(C_i) \alpha(C_j) \notag\\ &\quad \quad
{\cdot} \cos\!\big(\theta(C_i{+}\Delta C_i){-}\theta(C_j{+}\Delta C_j) {+} q_{i,j}(\varphi,\vartheta)\big)\Big] \\
& {=} \alpha(C_i) \alpha(C_j) \mathbb{E}_{\Delta C_i, \Delta C_j} \Big[\Re \Big\{e^{\jmath \theta(C_i{+}\Delta C_i)} \notag \\&  \kern77pt \cdot  e^{{-}\jmath \theta(C_j{+}\Delta C_j)} e^{\jmath q_{i,j}(\varphi,\vartheta)} \Big\}\Big] \\
& {=}\alpha(C_i) \alpha(C_j) \Re \Big\{ \! \mathbb{E}_{\Delta C_i} \! \big[\!e^{\jmath \theta(C_i{+}\Delta C_i)}\!\big] \notag \\ & \kern53pt \cdot  \mathbb{E}_{\Delta C_j} \big[e^{{-} \jmath \theta(C_j{+}\Delta C_j)}\big]  e^{\jmath q_{i,j}(\varphi,\vartheta)}\!\Big\} ,
\label{eq:intermediate_step_piecewise_model}
\end{align}
where~\eqref{eq:intermediate_step_piecewise_model} follows by interchanging the order of $\Re$ and $\mathbb{E}$, and exploiting that $\Delta C_i$, $\Delta C_j$ are i.i.d. random variables. Finally, \scalebox{1}{$G_{+}(C_i){=}\mathbb{E}_{\Delta C_i} \!\big[\!e^{\jmath \theta(C_i{+}\Delta C_i)}\!\big]$}, and \scalebox{1}{$G_{-}(C_j){=}\mathbb{E}_{\Delta C_j}\!\big[\!e^{{-} \jmath \theta(C_j{+}\Delta C_j)}\!\big]$} are calculated by~\eqref{eq:large_eq_cross_plus} and~\eqref{eq:large_eq_cross_minus}, given in Appendix~\ref{appendix:2}, respectively. Therefore, $\mathbb{E}_{\bm{\Delta c}}[\tilde{A}(\varphi,\vartheta)]$ can be obtained by substituting~\eqref{eq:large_eq_cross_plus} and~\eqref{eq:large_eq_cross_minus} in~\eqref{eq:intermediate_step_piecewise_model} and by using~\eqref{eq:intermediate_step_piecewise_model} %and~\eqref{eq:1st_term_mean_value}
in~\eqref{eq:mean_value_pattern_inti}, and is given by \textbf{Proposition~\ref{proposition:iid_gaussian_large_variations}}.
\begin{proposition}\label{proposition:iid_gaussian_large_variations}
The expectation of the RIS power radiation pattern, {$\tilde{A}(\varphi,\vartheta)$}, under the presence of large-variance i.i.d. Gaussian capacitance variations, is given by
\begin{align}
\mathbb{E}_{\bm{\Delta c}}&[\tilde{A}(\varphi,\vartheta)] {=}
\sum_{i=1}^N
\alpha^2(C_i) + \notag \\ \! & + 2 \sum_{i=1}^{N-1} \sum_{j{=}i{+}1}^N \Re\Big\{G_{+}(C_i)\, G_{-}(C_j)e^{\jmath q_{i,j}(\varphi,\vartheta)}\Big\},
\label{eq:mean_value_pattern_large_vars}
\end{align}
\begin{proof}
    The result follows directly from Section~\ref{subsec:Exp_value_high_variations}.
\end{proof}
\end{proposition}

\subsection{RIS Power Radiation Pattern Variance Derivation}
\label{sec:variance_computations}
Although the mean value of the amplitude $\tilde{A}(\varphi, \vartheta)$ provides the expected radiation behavior and serves as a solid baseline for optimizing the RIS configuration, deriving its variance is essential for evaluating the system's robustness. Specifically, the variance quantifies the uncertainty and the instantaneous fluctuations around the mean value, which are induced by the underlying controllable capacitance variations. By explicitly calculating the variance (or equivalently the standard deviation (STD)), the reliability of the proposed mean-based optimization approach can be justified. This guarantees that the actual, real-time performance of the RIS will closely track the expected theoretical pattern, ensuring that the system does not suffer from severe and unpredictable performance degradation. To this end, the variance of the RIS power radiation pattern w.r.t. the perturbations $\bm{\Delta c}$ is given by
\begin{align}
    \operatorname{Var} \big(\tilde{A}(\varphi,\vartheta)\big) 
    &= \mathbb{E}_{\bm{\Delta c}} \big[ \tilde{A}^2(\varphi,\vartheta) \big] - \mathbb{E}^{2}_{\bm{\Delta c}}[\tilde{A}(\varphi,\vartheta)],
    \label{eq:var_definition}
\end{align}
where $\mathbb{E}^{2}_{\bm{\Delta c}}[\tilde{A}(\varphi,\vartheta)]$ is given either by~\eqref{eq:mean_pattern_final} (small-variance variations) or by~\eqref{eq:mean_value_pattern_large_vars} (large-variance variations). For the first term in~\eqref{eq:var_definition}, the squared RIS power pattern is given by 
\begin{align}
\tilde{A}^2(\varphi,\vartheta)=&\sum_{i{=}1}^{N}\sum_{j{=}1}^{N}\sum_{l{=}1}^{N}\sum_{m{=}1}^{N} \alpha(C_i) \alpha(C_j) \alpha(C_l) \alpha(C_m)  \notag \\
    & \cdot \operatorname{exp} \Big( \jmath \big( \theta(C_i{+}\Delta C_i) - \theta(C_j{+}\Delta C_j) + q_{i,j} \notag \\
    & + \theta(C_l{+}\Delta C_l) - \theta(C_m{+}\Delta C_m) + q_{l,m} \big) \Big).\label{eq:squaredRIS_powerpattern}
\end{align}

To calculate the $\mathbb{E}_{\bm{\Delta c}}[\tilde{A}^2(\varphi,\vartheta)]$, the linearity property of the expectation operator is applied to~\eqref{eq:squaredRIS_powerpattern}. Since capacitance variations are statistically independent across RIS elements, multiplicative terms decouple depending on the indices $i, j, l$, and $m$. When indices coincide, the corresponding random variables cancel out or merge, altering the resulting expected value. Therefore, to accurately evaluate the mean value, the quadruple summation must be partitioned into five mutually exclusive subsets based on index equality. Defining the summand of~\eqref{eq:squaredRIS_powerpattern} as $\Psi_{i,j,l,m}$, the expected value is given by
\begin{align}
    \mathbb{E}_{\bm{\Delta c}}\big[\tilde{A}^2(\varphi,\vartheta)\big] = \sum_{k=1}^{5} E_k,
    \label{eq:mean_A_squared}
\end{align}
where each $E_k$ represents the sum of the expected values over a specific index combination subset, defined as follows:

\begin{itemize}
    \item \textbf{Subset~1 -- All indices are identical} ($1$~permutation): The indices satisfy $i{=}j{=}l{=}m$.
    \begin{align}
        E_1 = \sum_{i=1}^{N} \mathbb{E}_{\bm{\Delta c}}\big[ \Psi_{i,i,i,i} \big].
        \label{eq:var_comp_1}
    \end{align}
    \item \textbf{Subset~2 -- Three identical indices and one distinct} ($4$~permutations): e.g., $i{=}j{=}l {\neq} m$.
    \begin{align}
        E_2 {=}& \! \smash[b]{\mathop{\sum\sum}_{i \neq m} }\big( \mathbb{E}_{\bm{\Delta c}}\big[\Psi_{i,i,i,m}\big] + \mathbb{E}_{\bm{\Delta c}}\big[\Psi_{i,i,m,i}\big] \notag \\ & \quad \quad \quad +\mathbb{E}_{\bm{\Delta c}}\big[\Psi_{i,m,i,i}\big] + \mathbb{E}_{\bm{\Delta c}}\big[\Psi_{m,i,i,i}\big] \big).
        \label{eq:var_comp_2}
    \end{align}
    \item \textbf{Subset~3 -- Two distinct pairs of identical indices} ($3$~permutations): %The indices form two independent pairs (
    e.g., $i{=}j$ and $l{=}m$, with $i {\neq} l$.
    \begin{align}
       \!\!\!\!\!\! E_3 {=} \!\!\mathop{\sum  \! \sum}_{i \neq l}  \!\kern-1pt\Big( \! \mathbb{E}_{\bm{\Delta c}}\!\big[\! \Psi_{i,i,l,l}\! \big] {+}  \mathbb{E}_{\bm{\Delta c}}\!\big[\!\Psi_{i,l,i,l}\!\big] {+} \mathbb{E}_{\bm{\Delta c}}\!\big[\!\Psi_{i,l,l,i}\!\big] \! \Big)\!.
       \label{eq:var_comp_3}
    \end{align}
    \item \textbf{Subset~4 -- One pair of identical indices and two distinct} ($6$~permutations): Only two indices match, while the remaining two are unique (e.g., $i=j$, with $i \neq l \neq m$).
    \begin{align}
         E_4 {=}& \smash[b]{\mathop{\sum\sum\sum}_{i \neq l \neq m}} \big( 
        \mathbb{E}_{\bm{\Delta c}}\big[\Psi_{i,i,l,m}\big] + \mathbb{E}_{\bm{\Delta c}}\big[\Psi_{i,l,i,m}\big] \notag \\ & \quad  \quad \quad \quad  +\mathbb{E}_{\bm{\Delta c}}\big[\Psi_{i,l,m,i}\big] {+} \mathbb{E}_{\bm{\Delta c}}\big[\Psi_{l,i,i,m}\big] \notag \\ &  \quad \quad \quad \quad  +\mathbb{E}_{\bm{\Delta c}}\big[\Psi_{l,i,m,i}\big] {+} \mathbb{E}_{\bm{\Delta c}}\big[\Psi_{l,m,i,i}\big]
        \big).
        \label{eq:var_comp_4}
    \end{align}
     \item \textbf{Subset~5 -- All four indices are distinct} ($1$ permutation): The indices satisfy $i {\neq} j {\neq} l {\neq} m$. All random variables are independent, allowing the expectation to be fully factored.
    \begin{align}
        E_5 = \mathop{\sum\sum\sum\sum}_{i \neq j \neq l \neq m} \mathbb{E}_{\bm{\Delta c}}\big[ \Psi_{i,j,l,m} \big].
        \label{eq:var_comp_5}
    \end{align}
\end{itemize}
By evaluating the five components~\eqref{eq:var_comp_1}--\eqref{eq:var_comp_5} individually, the exact mean squared amplitude can be computed. To compute $\mathbb{E}_{\bm{\Delta \bm{c}}}[\Psi_{i,j,l,m}]$, deterministic components are isolated from the expectation operator. Introducing the deterministic amplitude-phase factor $D_{i,j,l,m}$ and the $k$th order moment of the random exponential term $\mu_k^{(x)}$, these base terms are defined as
\begin{align}
    D_{i,j,l,m} &{=} \alpha(C_i) \alpha(C_j) \alpha(C_l) \alpha(C_m) \operatorname{exp}(\jmath q_{i,j} {+} \jmath q_{l,m}),\\
    \mu_k^{(x)} {=} &\mathbb{E}\Big[\! \operatorname{exp}\!\big(\jmath k \cdot \theta(C_x {+} \Delta C_x)\big) \Big], \,\, k {\in} \{{-}2, {-}1, 1, 2\}.
    \label{eq:moments_variance}
\end{align}

Due to the mutual independence of capacitance variations across different elements, the expected values of the cross-products can be fully factored. The resulting expected values for all permutations across the five subsets are derived below:
\begin{itemize}
    \item \textbf{Subset 1 (All identical indices):} 
    \begin{align}
        \mathbb{E}_{\bm{\Delta c}}\big[\Psi_{i,i,i,i}\big] = D_{i,i,i,i}.
    \end{align}

    \item \textbf{Subset 2 (Three identical, one distinct):} 
    \begin{align}
            \mathbb{E}_{\bm{\Delta c}}\big[\Psi_{i,i,i,m}\big] &= D_{i,i,i,m} \, \mu_1^{(i)} \mu_{-1}^{(m)} \\
            \mathbb{E}_{\bm{\Delta c}}\big[\Psi_{i,i,m,i}\big] &= D_{i,i,m,i} \, \mu_{-1}^{(i)} \mu_1^{(m)} \\
            \mathbb{E}_{\bm{\Delta c}}\big[\Psi_{i,m,i,i}\big] &= D_{i,m,i,i} \, \mu_1^{(i)} \mu_{-1}^{(m)} \\
            \mathbb{E}_{\bm{\Delta c}}\big[\Psi_{m,i,i,i}\big] &= D_{m,i,i,i} \, \mu_{-1}^{(i)} \mu_1^{(m)}.
    \end{align}

    \item \textbf{Subset 3 (Two distinct pairs):} %Depending on the positions of the identical indices, the phase variations either cancel out completely or double.
        \begin{align}
            \mathbb{E}_{\bm{\Delta c}}\big[\Psi_{i,i,l,l}\big] &= D_{i,i,l,l} \\
            \mathbb{E}_{\bm{\Delta c}}\big[\Psi_{i,l,i,l}\big] &= D_{i,l,i,l} \, \mu_2^{(i)} \mu_{-2}^{(l)} \\
            \mathbb{E}_{\bm{\Delta c}}\big[\Psi_{i,l,l,i}\big] &= D_{i,l,l,i}.
        \end{align}

    \item \textbf{Subset 4 (One pair, two distinct):}
    \begin{align}
 \mathbb{E}_{\bm{\Delta c}}\big[\Psi_{i,i,l,m}\big] &= D_{i,i,l,m} \, \mu_1^{(l)} \mu_{-1}^{(m)} \\
            \mathbb{E}_{\bm{\Delta c}}\big[\Psi_{i,l,i,m}\big] &= D_{i,l,i,m} \, \mu_2^{(i)} \mu_{-1}^{(l)} \mu_{-1}^{(m)} \\
            \mathbb{E}_{\bm{\Delta c}}\big[\Psi_{i,l,m,i}\big] &= D_{i,l,m,i} \, \mu_{-1}^{(l)} \mu_1^{(m)} \\
            \mathbb{E}_{\bm{\Delta c}}\big[\Psi_{l,i,i,m}\big] &= D_{l,i,i,m} \, \mu_1^{(l)} \mu_{-1}^{(m)} \\
            \mathbb{E}_{\bm{\Delta c}}\big[\Psi_{l,i,m,i}\big] &= D_{l,i,m,i} \, \mu_{-2}^{(i)} \mu_1^{(l)} \mu_1^{(m)} \\
            \mathbb{E}_{\bm{\Delta c}}\big[\Psi_{l,m,i,i}\big] &= D_{l,m,i,i} \, \mu_1^{(l)} \mu_{-1}^{(m)}.
    \end{align}

    \item \textbf{Subset 5 (All distinct indices):} %All random variables are completely uncoupled.
    \begin{align}
        \mathbb{E}_{\bm{\Delta c}}\big[\Psi_{i,j,l,m}\big] = D_{i,j,l,m} \, \mu_1^{(i)} \mu_{-1}^{(j)} \mu_1^{(l)} \mu_{-1}^{(m)}.
    \end{align}
\end{itemize}

\subsubsection{Variance of the RIS Power Radiation Pattern, under small-variance Variations}\label{sec:var_small_var}
To explicitly compute the final expected values, for the small-variance additive i.i.d. Gaussian variations, the phase function is approximated as a linear function of the capacitance. Thus, the moment $\mu_k^{(x)}$ defined in~\eqref{eq:moments_variance} is analytically evaluated as follows
\begin{align}
        \mu_k^{(x)} = &\mathbb{E}_{\bm{\Delta c}}\big[ \operatorname{exp}\big(\jmath \, k \, \theta(C_x + \Delta C_x)\big) \big] \notag \\
        %{=}& \mathbb{E}_{\bm{\Delta c}}\Big[ \operatorname{exp}\big(\jmath \, k \, (a_2 C_x {+} b2 + a_2\Delta C_x)\big) \Big] \notag \\
        = &\operatorname{exp}(\jmath \, k \, (a_2 C_x {+} b2)) \, \mathbb{E}_{\bm{\Delta c}}\big[ \operatorname{exp}(\jmath k a_2 \Delta C_x) \big] \notag \\
        =& e^{\jmath \, k \, (a_2 C_x {+} b2)} \,\, e^{-\frac{1}{2}a^2_2k^2\sigma_c^2},
        \label{eq:moments_small_var}
\end{align}
which holds due to $\mathbb{E}_{\bm{\Delta c}}\!\big[ e^{\jmath k a_2 \Delta C_x} \big]{=}e^{-\frac{1}{2}a_2^2k^2\sigma_c^2}$ (proved in Appendix~\ref{appendix:1}). By substituting $k{\in}\{{\pm}2,{\pm}1\}$ into the general formula~\eqref{eq:moments_small_var}, the individual moment terms required for evaluating subsets 1--5 can be obtained. The derived terms, can then be substituted in~\eqref{eq:mean_A_squared} to yield the  $\mathbb{E}_{\bm{\Delta c}}\big[\tilde{A}^2(\varphi,\vartheta)\big]$.

\subsubsection{Variance of the RIS Power Radiation Pattern, under large-variance Variations}\label{sec:var_high_var} To explicitly compute the expected values, for the large-variance additive i.i.d. Gaussian variations case, the phase function is considered to be a piecewise linear approximation given by~\eqref{eq:piecewise_func} w.r.t. the capacitance. Thus, the general moment $\mu_k^{(x)}$ defined in~\eqref{eq:moments_variance} is evaluated as follows
\begin{align}
        \mu_k^{(x)} &= \mathbb{E}_{\bm{\Delta c}}\big[ \operatorname{exp}\big(\jmath \, k \, \theta(C_x + \Delta C_x)\big) \big] \notag \\
         %= & \int_{-\infty}^{+\infty} e^ {\jmath \, k \, \theta(C_x + \Delta C_x)} f(\Delta C_x) \text{d}\Delta C_x \notag \\ 
         &= e^{\jmath \, k \, (a_{21} C_x {+} b_{21})}\!\!\! \int_{-\infty}^{c_1-C_x}\!\!\!\!\! e^{\jmath k a_{21}\Delta C_x}f(\Delta C_x)\text{d}\Delta C_x \notag \\ 
          &\quad \,\, {+} e^{\jmath \, k \, (a_{22} C_x {+} b_{22})}\!\!\! \int_{c_1-C_x}^{c_2-C_x}\!\!\!\!\! e^{\jmath k a_{22}\Delta C_x}f(\Delta C_x)\text{d}\Delta C_x \notag \\ & \quad \,\, {+} e^{\jmath \, k \, (a_{23} C_x {+} b_{23})} \!\!\! \int_{c_2-C_x}^{+\infty}\!\!\!\!\! e^{\jmath k a_{23}\Delta C_x}f(\Delta C_x)\text{d}\Delta C_x  \\ {=} I_1 &e^{\jmath  k \, (a_{21} C_x {+} b_{21})} \!{+}\! I_2 e^{\jmath k \, (a_{22} C_x {+} b_{22})}  \!{+}\! I_3 e^{\jmath k \, (a_{23} C_x {+} b_{23})},
        \label{eq:moments_high_var}
\end{align}
where $f(\Delta C_x)$ is given by~\eqref{eq:gaussian_pdf_capac}, and $I_1$, $I_2$, $I_3$ are computed using~\eqref{eq:I1_integral},~\eqref{eq:I2_integral} and~\eqref{eq:I3_integral}, respectively and are given by %These integrals are computed using Mathematica (ver. 14.1) as follows: 
\begin{align}
    I_1 &= \frac{1}{2} e^{-\frac{1}{2} a_{21}^2 k^2 \sigma_c^2} \operatorname{erfc}\bigg( \frac{{-}c_1 {+} C_x {+} \jmath a_{21} k \sigma_c^2}{\sqrt{2} \sigma_c} \bigg) \label{eq:I1_integral}\\
    I_2 &= \frac{1}{2} e^{-\frac{1}{2} a_{22}^2 k^2 \sigma_c^2} \bigg( \operatorname{erf}\bigg( \frac{{-}c_1 {+} C_x {+} \jmath a_{22} k \sigma_c^2}{\sqrt{2} \sigma_c} \bigg) \notag \\ 
    & \quad \quad \quad \quad\quad \quad \quad-  \operatorname{erf}\bigg( \frac{{-}c_2 {+} C_x {+} \jmath a_{22} k \sigma_c^2}{\sqrt{2} \sigma_c} \bigg) \bigg) \label{eq:I2_integral} \\
    I_3 &= \frac{1}{2} e^{-\frac{1}{2} a_{23}^2 k^2 \sigma_c^2} \left( 1 {+} \operatorname{erf}\bigg( \frac{{-}c_2 {+} C_x {+} \jmath a_{23} k \sigma_c^2}{\sqrt{2} \sigma_c} \bigg) \right) \label{eq:I3_integral}.
\end{align}
By substituting $k{\in}\{{\pm}2,{\pm}1\}$ into the generalized formula~\eqref{eq:moments_high_var}, the individual moment terms required for evaluating subsets 1--5 can likewise be obtained for the large-variance case. The derived terms can then be
substituted in~\eqref{eq:mean_A_squared} to yield the exact expression for the mean squared amplitude $\mathbb{E}_{\bm{\Delta c}}\big[\tilde{A}^2(\varphi,\vartheta)\big]$.
\begin{figure*}[!tbp] 

    \centering
    
    \subfloat[$N=16 \times 16$, AoD: $(30^\circ, 0^\circ)$\label{fig:Fig_RIS_patterns_B_N2_lin}]{%
      \resizebox{0.32\textwidth}{!}
      %{\input{TikZ_Plots_IID_Lvar_noAmpl/Fig_RIS_Patterns_B_N2_lin}}
      {\includegraphics[width=1\textwidth]{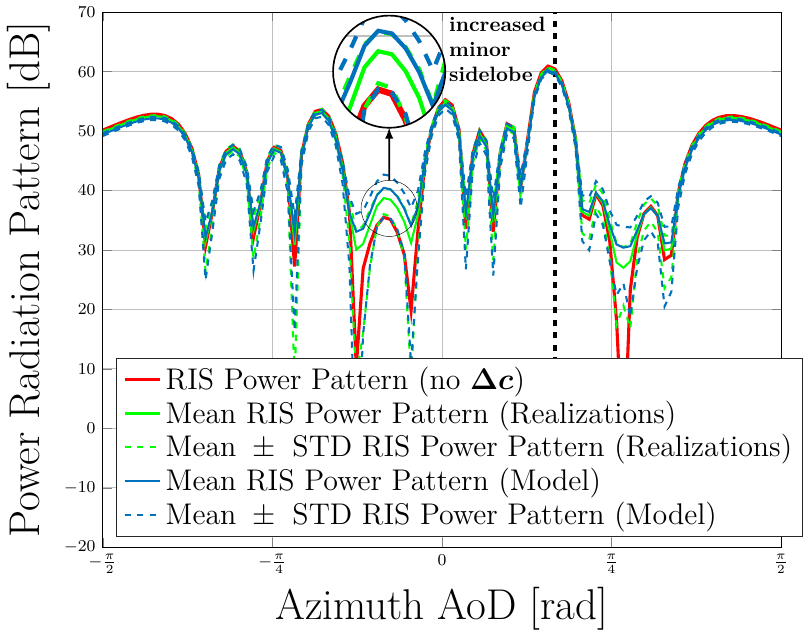}}
    }  
    \subfloat[$N=32 \times 32$, AoD: $(30^\circ, 0^\circ)$\label{fig:Fig_RIS_patterns_B_N3_lin}]{
    \resizebox{0.32\textwidth}{!}
      %{\input{TikZ_Plots_IID_Lvar_noAmpl/Fig_RIS_Patterns_B_N3_lin}}
      {\includegraphics[width=1\textwidth]{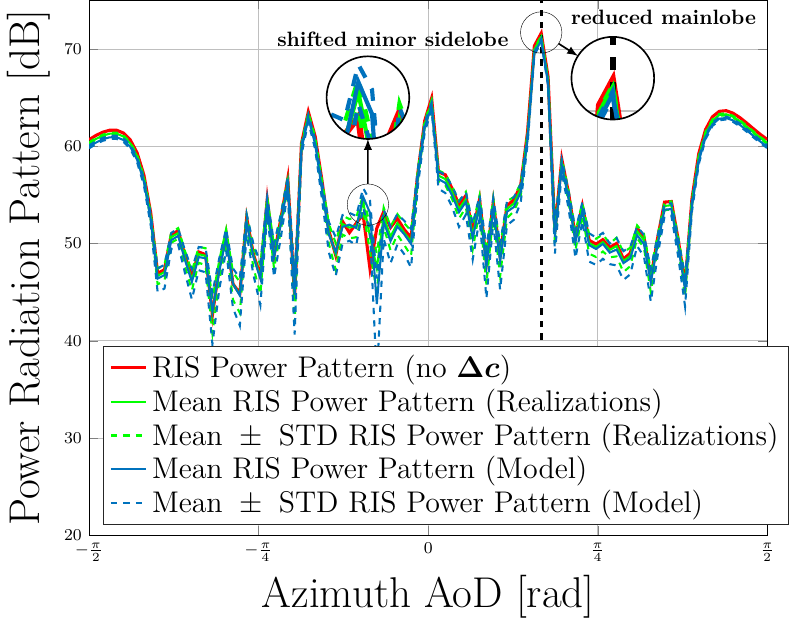}}
    }
    \subfloat[$N=64 \times64$, AoD: $(30^\circ, 0^\circ)$\label{fig:Fig_RIS_patterns_B_N4_lin}]{
    \resizebox{0.32\textwidth}{!}
      %{\input{TikZ_Plots_IID_Lvar_noAmpl/Fig_RIS_Patterns_B_N4_lin}}
      {\includegraphics[width=1\textwidth]{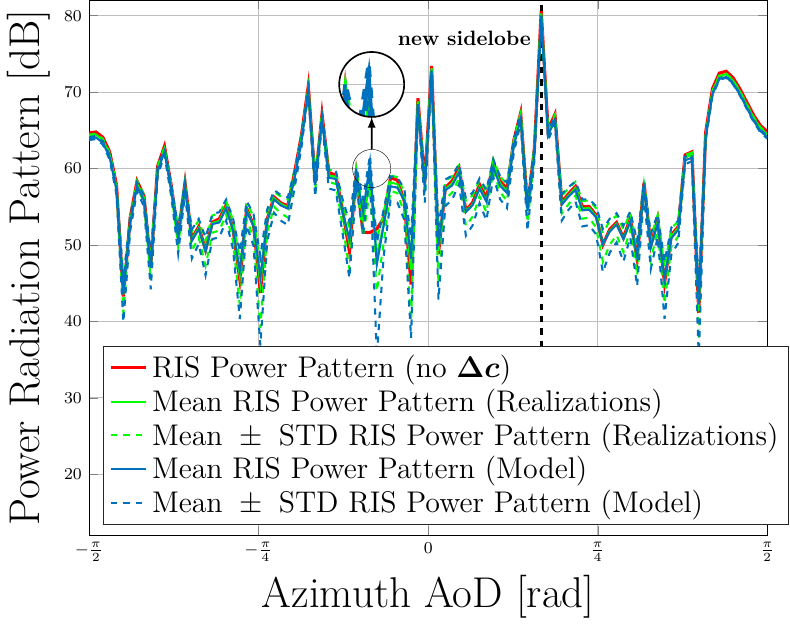}}
    }
    \\
    % \subfloat[$N=8 \times 8$\label{fig:Fig_RIS_patterns_B_N1_lin2}]{%
    % \resizebox{0.24\textwidth}{!}
    %   {\input{TikZ_Plots_IID_Lvar_noAmpl/Fig_RIS_Patterns_B_N1_lin2}} 
    % } 
    \subfloat[$N=16 \times 16$, AoD: $(15^\circ, 0^\circ)$\label{fig:Fig_RIS_patterns_B_N2_lin2}]{%
    \resizebox{0.32\textwidth}{!}
      %{\input{TikZ_Plots_IID_Lvar_noAmpl/Fig_RIS_Patterns_B_N2_lin2}}
      {\includegraphics[width=1\textwidth]{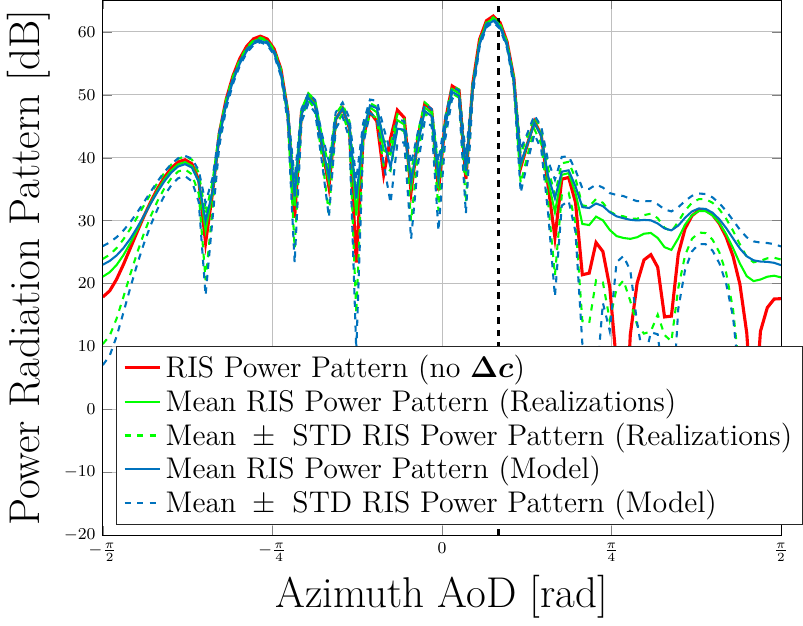}}
    }  
    \subfloat[$N=32 \times 32$, AoD: $(15^\circ, 0^\circ)$\label{fig:Fig_RIS_patterns_B_N3_lin2}]{
    \resizebox{0.32\textwidth}{!}
      %{\input{TikZ_Plots_IID_Lvar_noAmpl/Fig_RIS_Patterns_B_N3_lin2}}
      {\includegraphics[width=1\textwidth]{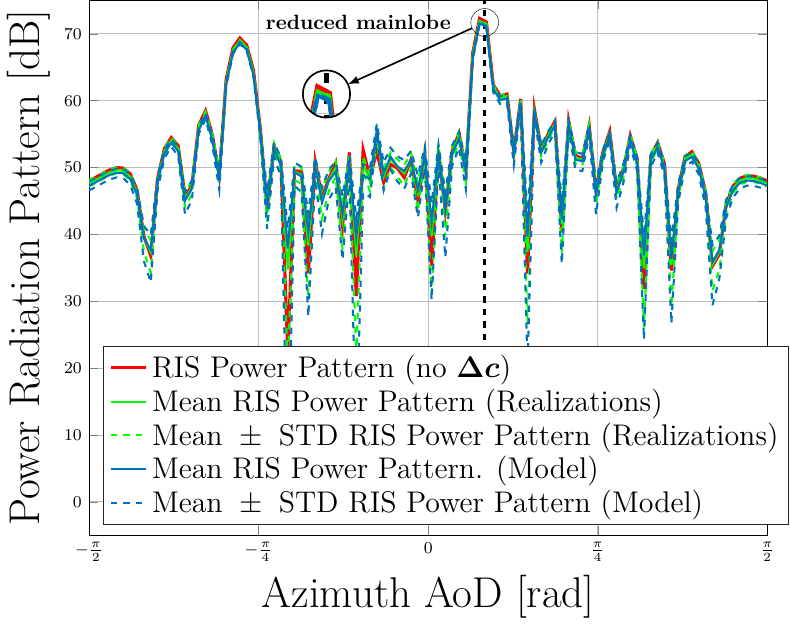}}
    } 
    \subfloat[$N=64 \times64$, AoD: $(15^\circ, 0^\circ)$\label{fig:Fig_RIS_patterns_B_N4_lin2}]{
    \resizebox{0.32\textwidth}{!}
        %{\input{TikZ_Plots_IID_Lvar_noAmpl/Fig_RIS_Patterns_B_N4_lin2}}
        {\includegraphics[width=1\textwidth]{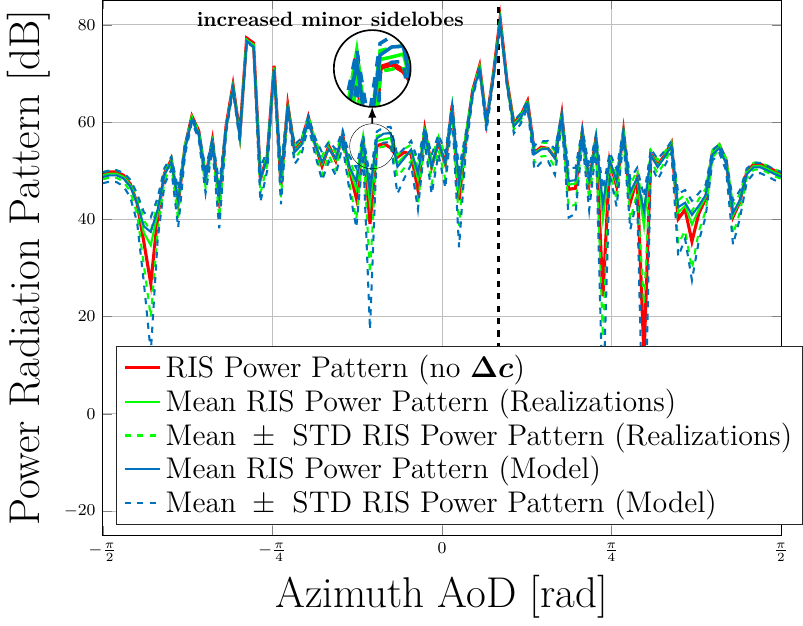}}
    }    
   \caption{Impact of RIS-elements' capacitance variations on the RIS radiation pattern under the linear TLM model (low variability). The RIS is illuminated from the azimuth-elevation angle pair $(-15^\circ, 0^\circ)$ and steered towards $(30^\circ, 0^\circ)$ (top row) and $(15^\circ, 0^\circ)$ (bottom row), with the target AoD indicated by the gray dashed line. The RIS configuration is 1-bit quantized and optimized via~\eqref{eq:MRT_solution}. Curves represent: the nominal RIS power pattern without variations (\textbf{red}); the empirical mean RIS pattern ($\pm$ STD) from $10000$ Monte Carlo runs under i.i.d. Gaussian variations with $\sigma_c{=}C_{\text{off}}/25$ (\textbf{green}); and the theoretical mean RIS pattern ($\pm$ STD) as per Section~\ref{subsec:Exp_value_iid} (\textbf{blue}). Results are sampled at 100 AoDs for each of three RIS sizes, $16{\times}16$, $32{\times}32$, $64{\times}64$.}
\label{fig:Fig_RIS_patterns_varying_N_lin_model}
\end{figure*}

%77
\section{RIS Design under Capacitance Variations}
\label{sec:RIS_optimization}
To maximize the RIS radiated power at a target angle $(\varphi_0, \vartheta_0)$ under infinite angular resolution, and for the variations-free case, the optimal RIS configuration is derived by maximizing the inner product in~\eqref{eq:power_factor}. This yields the maximum ratio transmission (MRT) solution~\cite{765552}
\begin{equation}
    \bm{\omega}_{\text{opt}} = (\mathbf{a}_{\text{RIS}}(\varphi_{\text{AoA}},\vartheta_{\text{AoA}}) \odot \mathbf{a}^{*}_{\text{RIS}}(\varphi, \vartheta))^{*}.
    \label{eq:MRT_solution}
\end{equation} 
While $\bm{\omega}_{\text{opt}}$ serves as the ideal continuous-domain benchmark, actual hardware realizations inevitably deviate from this nominal baseline. Crucially, even if high-precision phase quantization is employed to approximate this optimum, physical capacitance variations ($\Delta C_i, \forall i$) inherently persist at the element level. Consequently, these intrinsic variations inherently degrade the RIS power pattern, as demonstrated in Section~\ref{sec:simulation_results}; thus, this detrimental impact must be explicitly accounted for during the RIS optimization for radiation pattern shaping.

Utilizing the variation-aware models of Section~\ref{sec:model_formulation}, the final objective of this work, is to find the RIS capacitance configuration, $\bm{c} {=} [C_1, C_2, {\ldots}, C_N]^T$, that maximizes the expected reflected power towards a target angle $(\varphi_0, \vartheta_0)$, $\mathbb{E}_{\bm{\Delta c}}[\tilde{A}(\varphi_0,\vartheta_0)]$. Accordingly, the optimization problem is formulated as
\begin{equation}\label{eq:opt_problem}
\begin{aligned}
& \smash[b]{\underset{\bm{c}}{\operatorname{maximize}}} \quad
\mathbb{E}_{\bm{\Delta c}}[\tilde{A}(\varphi_0,\vartheta_0)], \\ &\quad \quad\quad C_i \in \{C_{\text{on}}, C_{\text{off}}\}, \quad \forall i = 1, 2, \ldots, N.
\end{aligned}
\end{equation} 
Due to the discrete nature of the available capacitance states, an exhaustive search would require the evaluation of $2^N$ possible configurations, which is computationally prohibitive for realistically sized arrays. Consequently, a low complexity greedy algorithm is adopted to solve~\eqref{eq:opt_problem}. The proposed greedy approach, show in Algorithm~\ref{alg:greedy_opt}, iteratively optimizes the RIS configuration element by element. Starting from an initial setup, the algorithm sequentially toggles each element's state between $C_{\text{on}}$ and $C_{\text{off}}$, evaluating the resulting mean radiated power at the target angle. A state change is retained only if it increases the target power; otherwise, it is reverted. This process repeats until all elements are evaluated.
\begin{algorithm}[t]
\KwIn{Target angle $(\varphi_0, \vartheta_0)$, initial configuration $\mathbf{c}$, selected variations model: Sections \ref{subsec:Exp_value_iid}--\ref{subsec:Exp_value_high_variations}}
\KwOut{The optimized capacitance configuration $\bm{c}_{\text{opt}}$}
\nl $P_{\max} = \mathbb{E}_{\bm{\Delta c}}[\tilde{A}(\varphi_0,\vartheta_0;\bm{c})]$\\
\nl \For {$i=1,\ldots,N$}{
\nl    $\bm{c}_{\text{old}} = \bm{c}_i$\\
\nl    $\bm{c}_i \gets$ Toggle state between $C_{\text{on}}$ and $C_{\text{off}}$\\
\nl    $P_{\text{new}} = \mathbb{E}_{\bm{\Delta c}}[\tilde{A}(\varphi_0,\vartheta_0;\bm{c})]$\\
\nl  \textbf{if} $P_{\text{new}} > P_{\max}$ \textbf{then} $P_{\max} {=} P_{\text{new}}$ \textbf{else} $\bm{c}_i {=} \bm{c}_{\text{old}}$
% \nl    \eIf{$P_{\text{new}} > P_{\max}$}{
%        $P_{\max} = P_{\text{new}}$
%        }{
%        $\bm{c}_i = \bm{c}_{\text{old}}$
%        }
}
\nl Obtain $\bm{c}_{\text{opt}} = \bm{c}$
\caption{Robust RIS Radiation Pattern Design} \label{alg:greedy_opt}
\end{algorithm}
The expected power $\mathbb{E}_{\bm{\Delta c}}[\tilde{A}(\varphi_0,\vartheta_0)]$ is evaluated using the analytical models in Section~\ref{sec:model_formulation} and the selected formulation, exploited into Algorithm~\ref{alg:greedy_opt}, depends on the nature of i.i.d. variations, accounting either small or large variations case.
\begin{figure*}[t!] 

    \centering
    % \subfloat[$N=8 \times 8$\label{fig:Fig_RIS_patterns_D_N1_plin}]{
    % \resizebox{0.24\textwidth}{!}
    % {\input{Tikz_Plots_IID_MODEL_Hvar/Fig_RIS_Patterns_D_N1_plin}}
    % }
    \subfloat[$N=16 \times 16$, AoD: $(30^\circ, 0^\circ)$\label{fig:Fig_RIS_patterns_D_N2_plin}]{%
    \resizebox{0.32\textwidth}{!}
      %{\input{Tikz_Plots_IID_MODEL_Hvar/Fig_RIS_Patterns_D_N2_plin}}
      {\includegraphics[width=\textwidth]{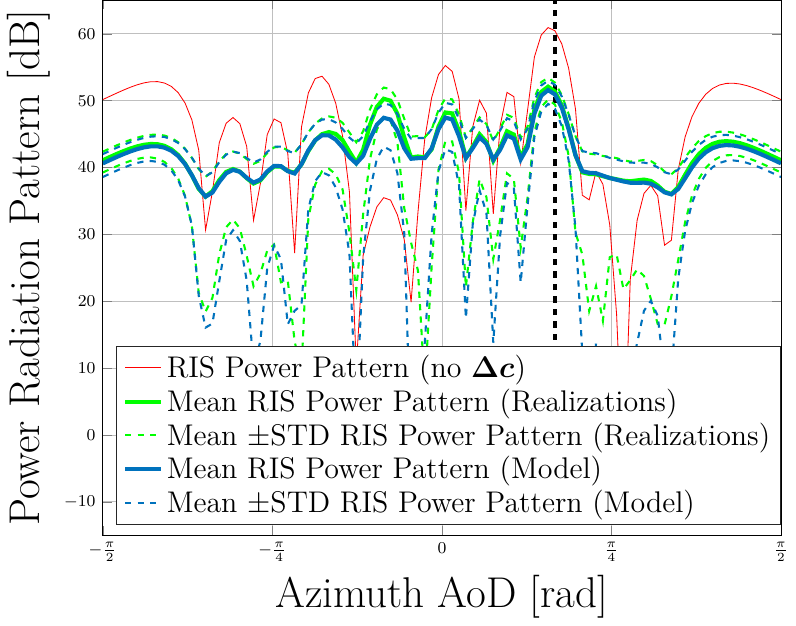}}
    } 
    \subfloat[$N=32 \times 32$, AoD: $(30^\circ, 0^\circ)$\label{fig:Fig_RIS_patterns_D_N3_plin}]{
    \resizebox{0.32\textwidth}{!}
      %{\input{Tikz_Plots_IID_MODEL_Hvar/Fig_RIS_Patterns_D_N3_plin}}
      {\includegraphics[width=\textwidth]{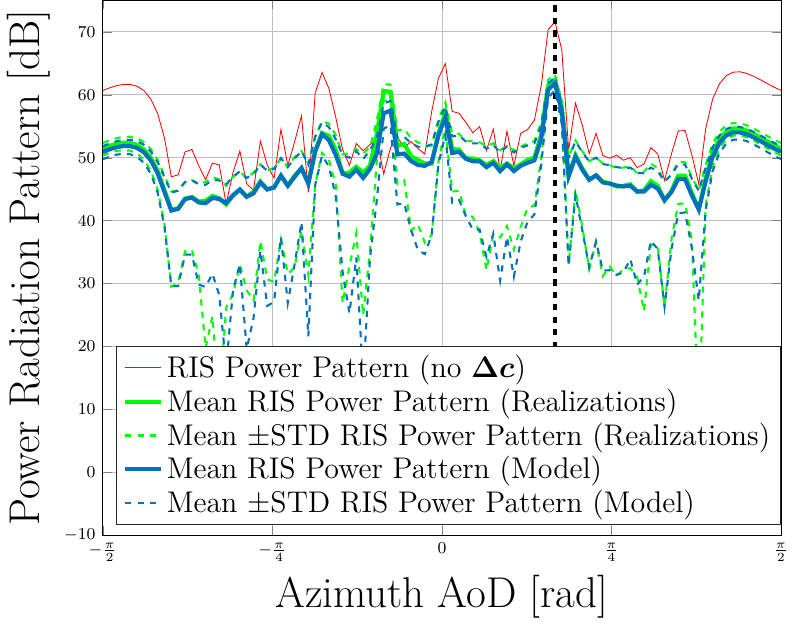}}
    }
    \subfloat[$N=64 \times 64$, AoD: $(30^\circ, 0^\circ)$\label{fig:Fig_RIS_patterns_D_N4_plin}]{%
    \resizebox{0.32\textwidth}{!}
      %{\input{Tikz_Plots_IID_MODEL_Hvar/Fig_RIS_Patterns_D_N4_plin}}
      {\includegraphics[width=\textwidth]{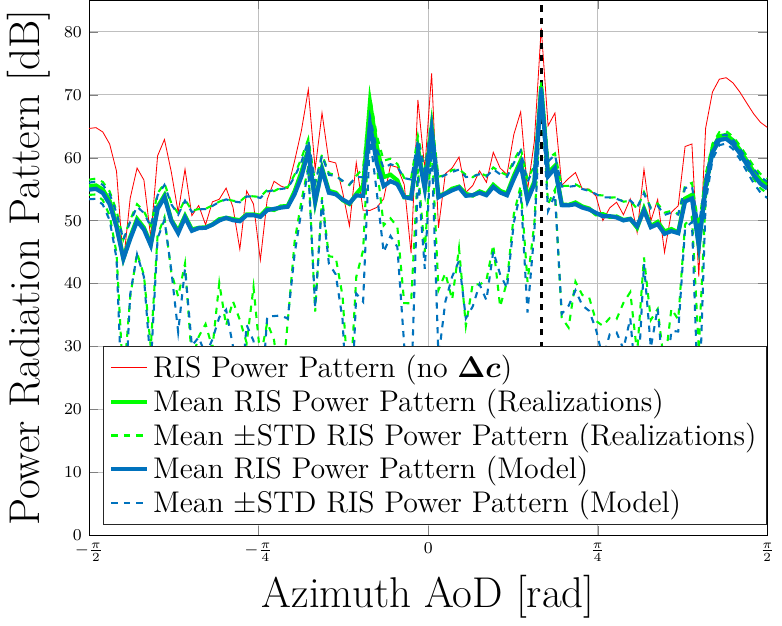}}
    }
    \\
    % \subfloat[$N=8 \times 8$\label{fig:Fig_RIS_patterns_D_N1_plin2}]{
    % \resizebox{0.24\textwidth}{!}
    % {\input{Tikz_Plots_IID_MODEL_Hvar/Fig_RIS_Patterns_D_N1_plin2}}
    % }
    \subfloat[$N=16 \times 16$, AoD: $(15^\circ, 0^\circ)$\label{fig:Fig_RIS_patterns_D_N2_plin2}]{%
    \resizebox{0.32\textwidth}{!}
      %{\input{Tikz_Plots_IID_MODEL_Hvar/Fig_RIS_Patterns_D_N2_plin2}}
      {\includegraphics[width=\textwidth]{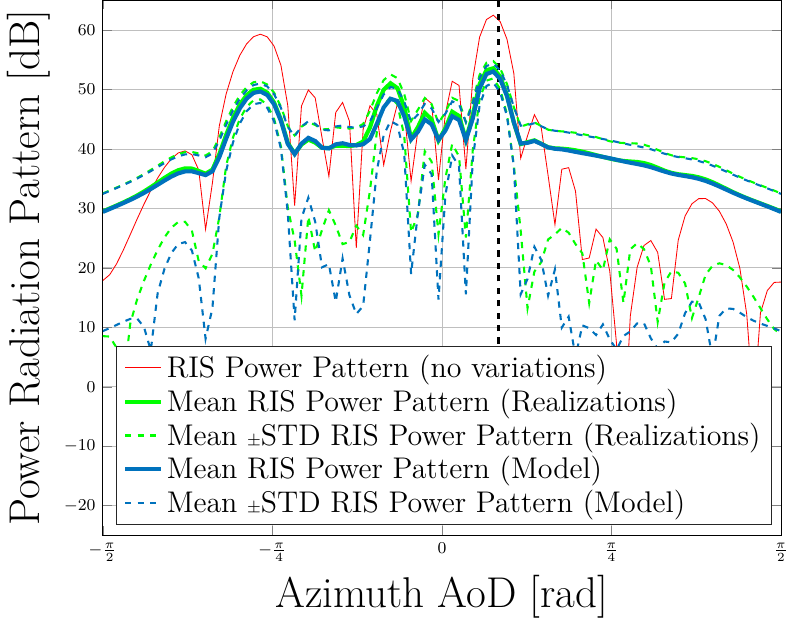}}
    } 
    \subfloat[$N=32 \times 32$, AoD: $(15^\circ, 0^\circ)$\label{fig:Fig_RIS_patterns_D_N3_plin2}]{
    \resizebox{0.32\textwidth}{!}
      %{\input{Tikz_Plots_IID_MODEL_Hvar/Fig_RIS_Patterns_D_N3_plin2}}
      {\includegraphics[width=\textwidth]{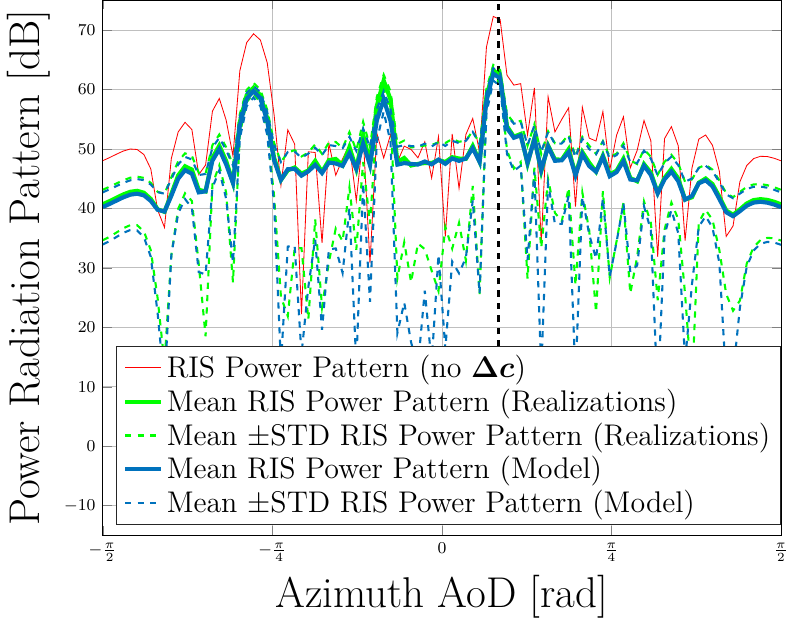}}
    }
    \subfloat[$N=64 \times 64$, AoD: $(15^\circ, 0^\circ)$\label{fig:Fig_RIS_patterns_D_N4_plin2}]{%
    \resizebox{0.32\textwidth}{!}
      %{\input{Tikz_Plots_IID_MODEL_Hvar/Fig_RIS_Patterns_D_N4_plin2}}
      {\includegraphics[width=\textwidth]{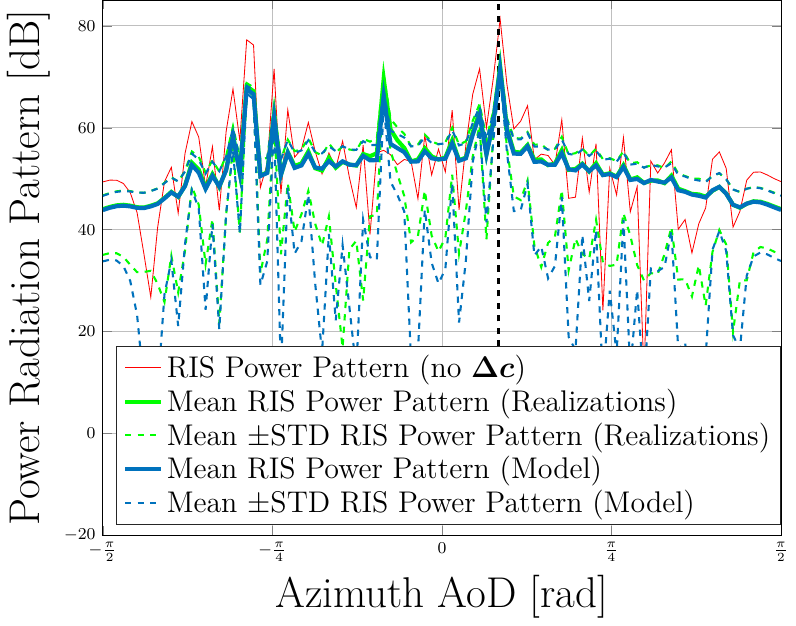}}
    }
%\caption{Impact of RIS-elements' capacitance variations on the RIS power radiation pattern. The RIS is illuminated from the azimuth-elevation angle pair \textbf{$({-}15^\circ, 0^\circ)$} and is configured to steer the reflected signal towards $(30^\circ, 0^\circ)$ (1st row Figs.) and $(15^\circ, 0^\circ)$ (2nd row Figs.). The targeted azimuth AoD is depicted with a gray dashed line. The RIS configuration is optimized according to~\eqref{eq:MRT_solution} and quantized for a $1$-bit implementation. The \textbf{red curve} depicts the (1-bit) quantized nominal RIS power radiation pattern when no accounting for per-element capacitance variations in the optimization. The \textbf{green curve} represents the mean (and $\pm$ STD) RIS power radiation pattern (of the same RIS configuration) under hardware imperfections, computed by averaging $10000$ Monte Carlo realizations ($1000$ vectors across $10$ random seeds), where the capacitance variation follows an i.i.d. Gaussian distribution with parameters $(\bm{\mu}, \bm{\Sigma}){=}(\bm{0},\sigma_c^{2}\bm{I})$ as given in~\eqref{eq:gaussian_pdf_capac}, with $\sigma_c{=}C_{\text{off}}/4$. The \textbf{blue curve} corresponds to the theoretical calculation of the mean (and $\pm$ STD) RIS power patterns based on the model described in Section~\ref{subsec:Exp_value_high_variations}. All patterns are sampled at $100$ distinct AoD, and the results are presented for varying numbers of RIS elements. This case corresponds to the piecewise-linear model of TLM, concerning high per-element capacitance variability.\vspace{0.1cm}}
\caption{Impact of capacitance variations on the RIS radiation pattern under the piecewise-linear TLM model (high variability). The RIS is illuminated from the azimuth-elevation angle pair $(-15^\circ, 0^\circ)$ and steered towards $(30^\circ, 0^\circ)$ (top row) and $(15^\circ, 0^\circ)$ (bottom row), with the target AoD indicated by the gray dashed line. Configurations are 1-bit quantized and optimized via~\eqref{eq:MRT_solution}. Curves represent: the nominal RIS power pattern without variations (\textbf{red}); the empirical mean RIS pattern ($\pm$ STD) from $10000$ Monte Carlo runs under i.i.d. Gaussian variations with $\sigma_c{=}C_{\text{off}}/4$ (\textbf{green}); and the theoretical mean RIS pattern ($\pm$ STD) as per Section~\ref{subsec:Exp_value_high_variations} (\textbf{blue}). Results are sampled at 100 AoDs for each of three RIS sizes $16{\times}16$, $32{\times}32$, $64{\times}64$.}
\label{fig:Fig_RIS_patterns_varying_N_plin_model}
\end{figure*}

\section{Simulation Results and Discussion}
\label{sec:simulation_results}
This section presents simulation results, which illustrate the impact of both small-variance (Fig.~\ref{fig:Fig_RIS_patterns_varying_N_lin_model}) and large-variance (Fig.~\ref{fig:Fig_RIS_patterns_varying_N_plin_model}) capacitance variations on the RIS power radiation pattern. Specifically, two scenarios are investigated: one in which the RIS is configured to steer the reflected signal towards an anomalous angle of $30^\circ$, and another in which the signal is reflected towards the specular direction of $15^\circ$. In both cases (Figs.~\ref{fig:Fig_RIS_patterns_varying_N_lin_model} and~\ref{fig:Fig_RIS_patterns_varying_N_plin_model}), the performance is evaluated across different RIS sizes, and it is shown that the beamforming gain is enhanced as the number of RIS elements increases, a fact that is already known in the antenna array processing theory.

\begin{figure*}[t!] 
    \centering
    \subfloat[$N=16 \times 16$, AoD: $(30^\circ, 0^\circ)$\label{fig:Fig_RIS_patterns30_N1_lin}]{%
    \resizebox{0.33\textwidth}{!}
      %{\input{Model_validation_low_var/model_val_case30_N1}}
      {\includegraphics[width=\textwidth]{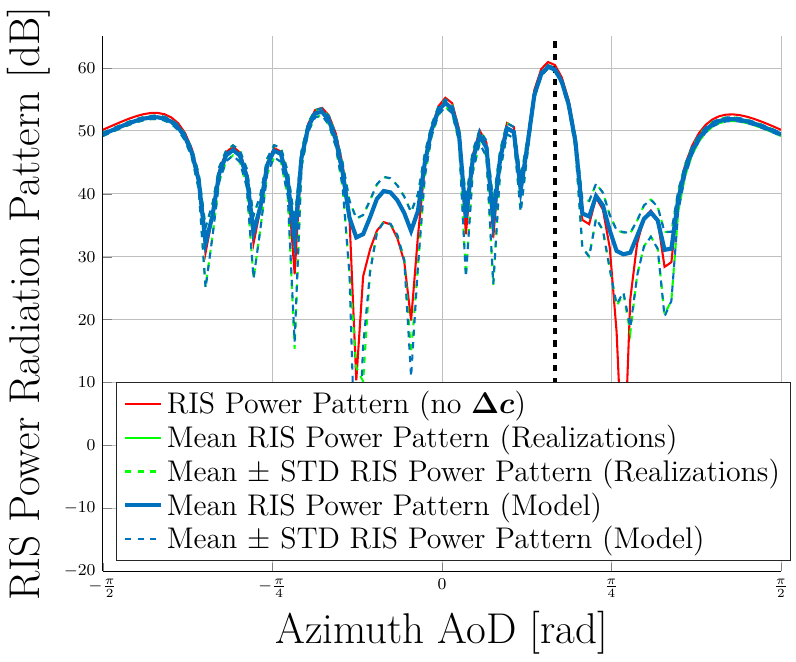}}
    } 
    \subfloat[$N=32 \times 32$, AoD: $(30^\circ, 0^\circ)$\label{fig:Fig_RIS_patterns30_N2_lin}]{
    \resizebox{0.325\textwidth}{!}
      %{\input{Model_validation_low_var/model_val_case30_N2}}
      {\includegraphics[width=\textwidth]{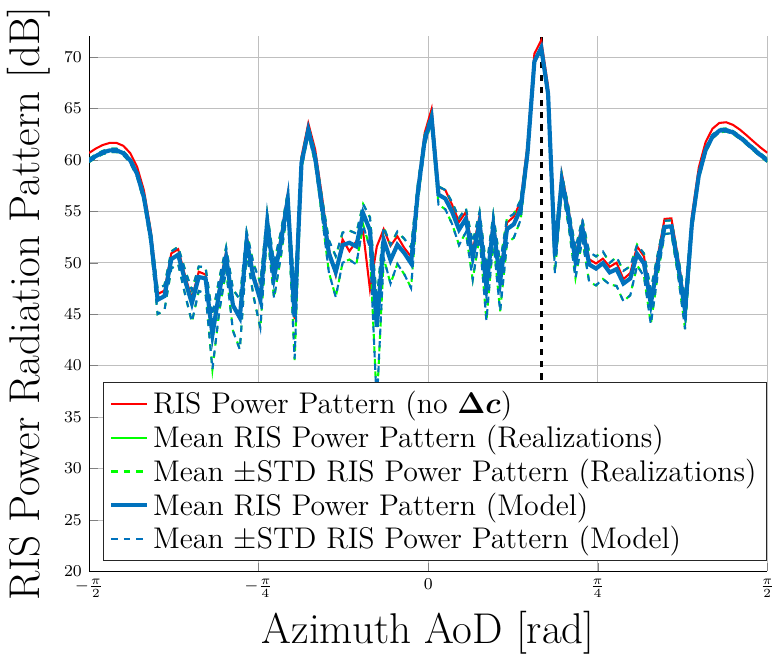}}
    }
    \subfloat[$N=64 \times 64$, AoD: $(30^\circ, 0^\circ)$\label{fig:Fig_RIS_patterns30_N3_lin}]{%
    \resizebox{0.325\textwidth}{!}
      %{\input{Model_validation_low_var/model_val_case30_N3}}
      {\includegraphics[width=\textwidth]{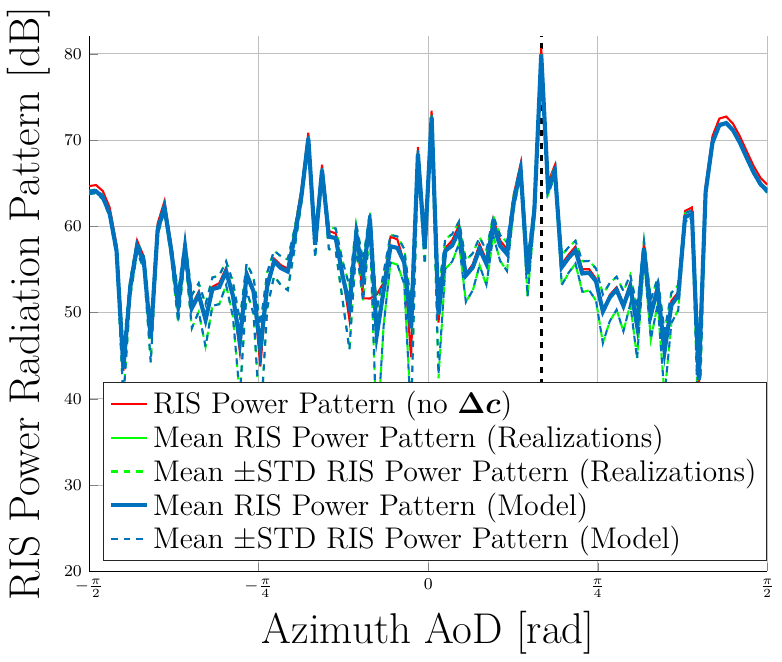}}
    }
    \\
    \subfloat[$N=16 \times 16$, AoD: $(15^\circ, 0^\circ)$\label{fig:Fig_RIS_patterns15_N1_lin}]{%
    \resizebox{0.32\textwidth}{!}
      %{\input{Model_validation_low_var/model_val_case15_N1}}
      {\includegraphics[width=\textwidth]{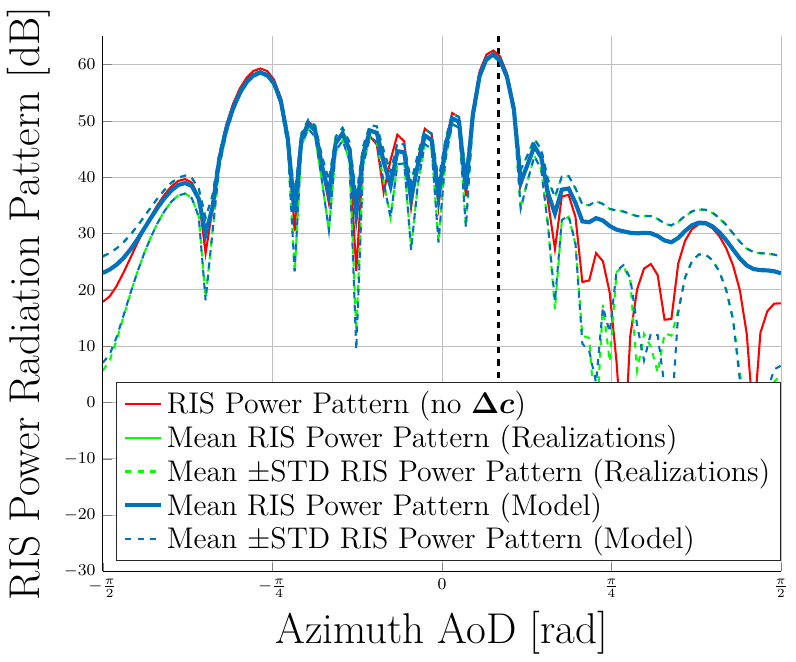}}
    } 
    \subfloat[$N=32 \times 32$, AoD: $(15^\circ, 0^\circ)$\label{fig:Fig_RIS_patterns15_N2_lin}]{
    \resizebox{0.32\textwidth}{!}
      %{\input{Model_validation_low_var/model_val_case15_N2}}
      {\includegraphics[width=\textwidth]{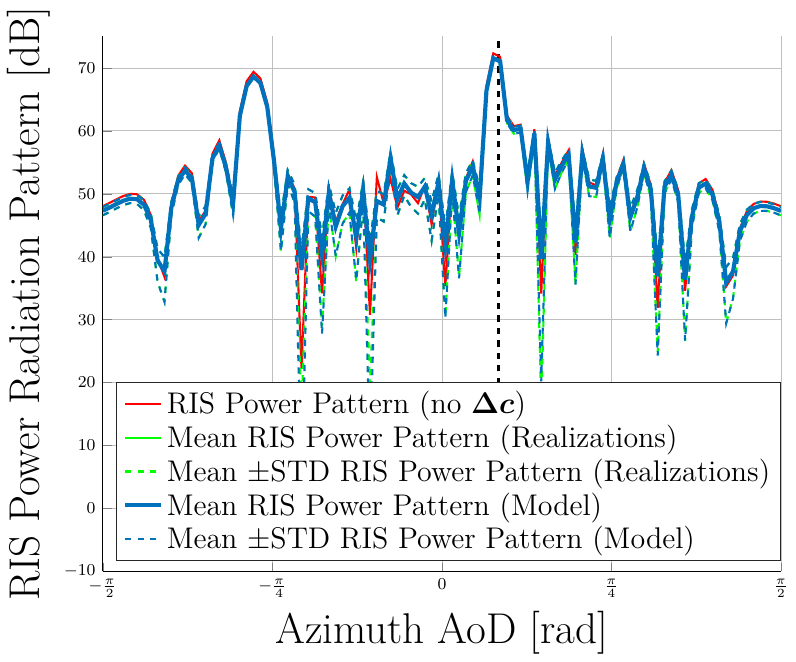}}
    }
    \subfloat[$N=64 \times 64$, AoD: $(15^\circ, 0^\circ)$\label{fig:Fig_RIS_patterns15_N3_lin}]{%
    \resizebox{0.32\textwidth}{!}
      %{\input{Model_validation_low_var/model_val_case15_N3}}
      {\includegraphics[width=\textwidth]{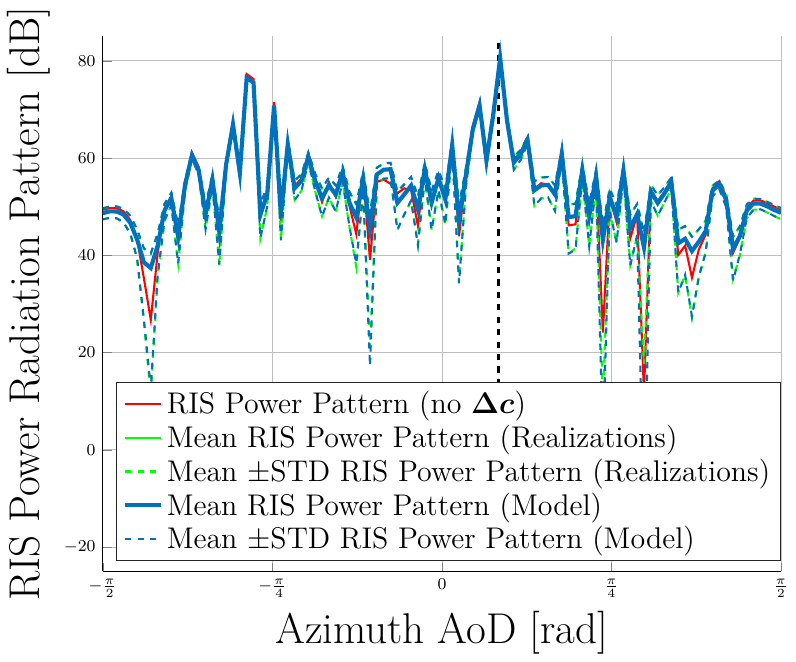}}
    }
\caption{Validation of capacitance variations impact on the RIS radiation pattern under the linear model (low variability) assumption. This case supposes that both realizations and the model (Section~\ref{subsec:Exp_value_iid}) assume nominal amplitude and linear phase for the RIS element reflection coefficients. The RIS is illuminated from $(-15^\circ, 0^\circ)$ and steered towards $(30^\circ, 0^\circ)$ (top row) and $(15^\circ, 0^\circ)$ (bottom row), with the target AoD indicated by the gray dashed line. The computed RIS configuration is 1-bit quantized and optimized via~\eqref{eq:MRT_solution}. Curves represent: the nominal pattern without variations (\textbf{red}); the empirical mean RIS power pattern ($\pm$ STD) from $10000$ Monte Carlo runs under i.i.d. Gaussian variations with $\sigma_c{=}C_{\text{off}}/25$ (\textbf{green}); and the theoretical mean RIS power pattern ($\pm$ STD) as per Section~\ref{subsec:Exp_value_iid} (\textbf{blue}). Results are sampled at 100 AoDs for each of three RIS sizes $16{\times}16$, $32{\times}32$, $64{\times}64$.}
\label{fig:Fig_RIS_patterns_varying_N_lin_model_valid}
\end{figure*}

\begin{figure*}[t!] 

    \centering
    \subfloat[$N=16 \times 16$, AoD: $(30^\circ, 0^\circ)$ \label{fig:Fig_RIS_patterns30_N1_plin}]{%
    \resizebox{0.317\textwidth}{!}
      %{\input{Model_validation_high_var/model_val_case30_N1}}
      {\includegraphics[width=\textwidth]{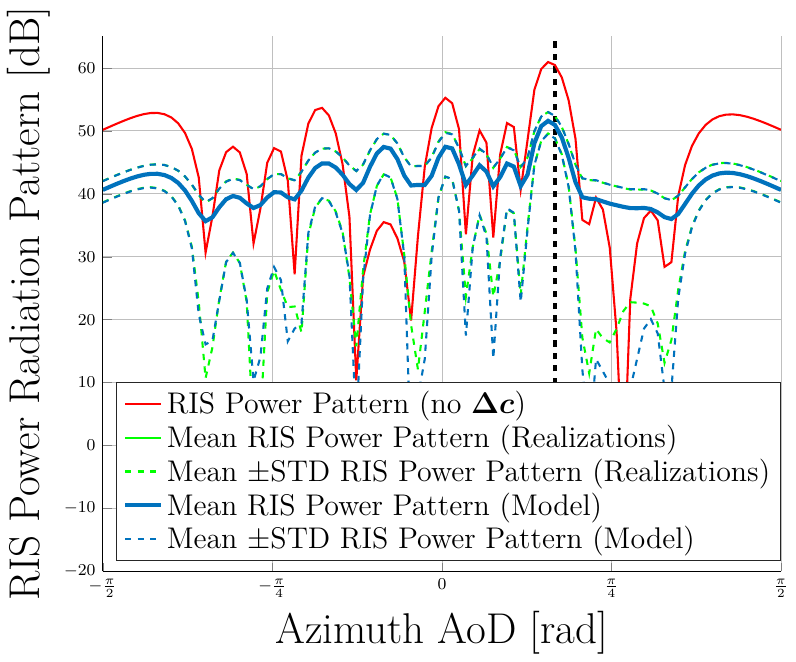}}
    } 
    \subfloat[$N=32 \times 32$, AoD: $(30^\circ, 0^\circ)$\label{fig:Fig_RIS_patterns30_N2_plin}]{
    \resizebox{0.317\textwidth}{!}
      %{\input{Model_validation_high_var/model_val_case30_N2}}
      {\includegraphics[width=\textwidth]{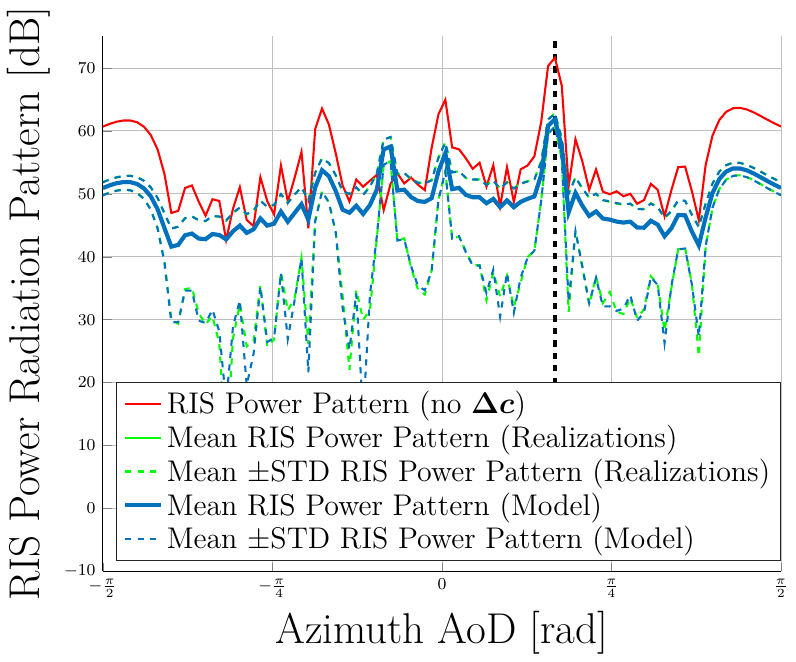}}
    }
    \subfloat[$N=64 \times 64$, AoD: $(30^\circ, 0^\circ)$\label{fig:Fig_RIS_patterns30_N3_plin}]{%
    \resizebox{0.317\textwidth}{!}
      %{\input{Model_validation_high_var/model_val_case30_N3}}
      {\includegraphics[width=\textwidth]{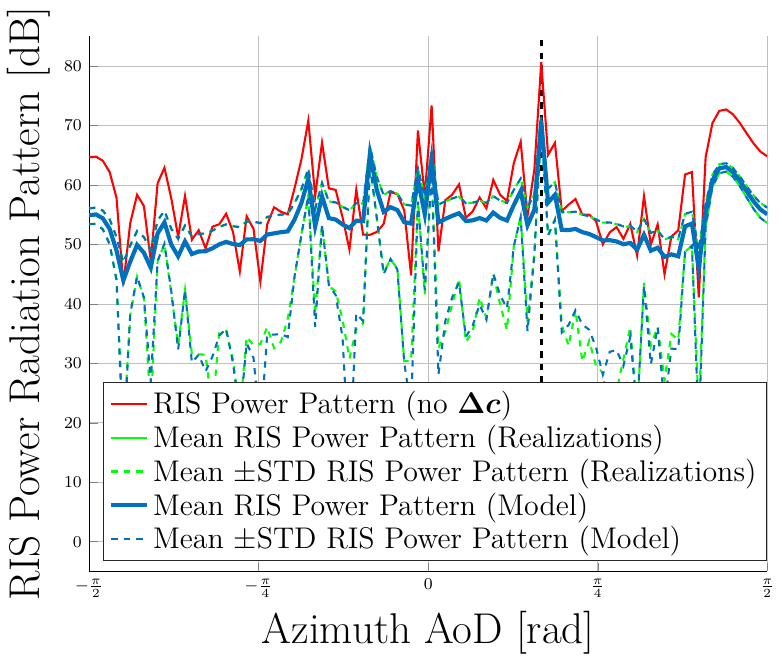}}
    }
    \\
    \subfloat[$N=16 \times 16$, AoD: $(15^\circ, 0^\circ)$\label{fig:Fig_RIS_patterns15_N1_plin}]{%
    \resizebox{0.317\textwidth}{!}
      %{\input{Model_validation_high_var/model_val_case15_N1}}
      {\includegraphics[width=\textwidth]{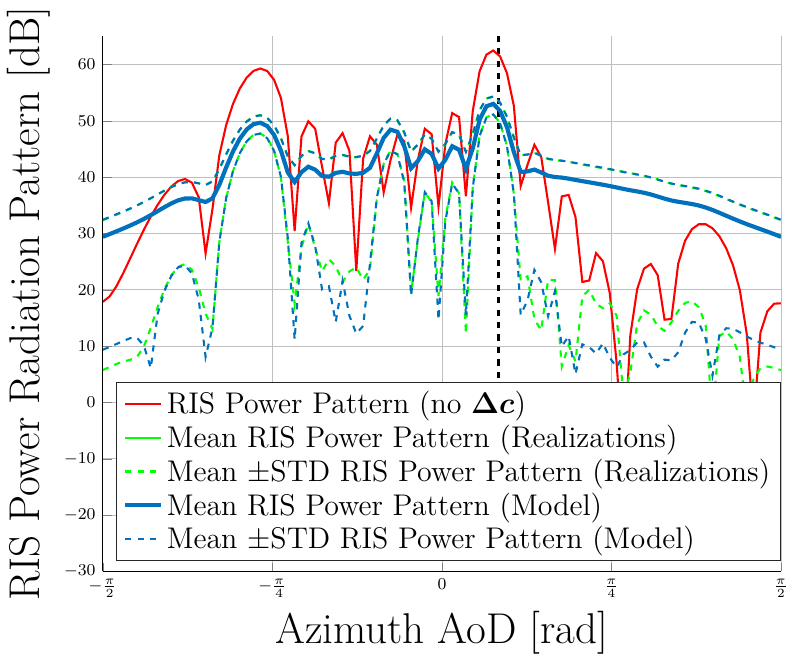}}
    } 
    \subfloat[$N=32 \times 32$, AoD: $(15^\circ, 0^\circ)$\label{fig:Fig_RIS_patterns15_N2_plin}]{
    \resizebox{0.317\textwidth}{!}
      %{\input{Model_validation_high_var/model_val_case15_N2}}
      {\includegraphics[width=\textwidth]{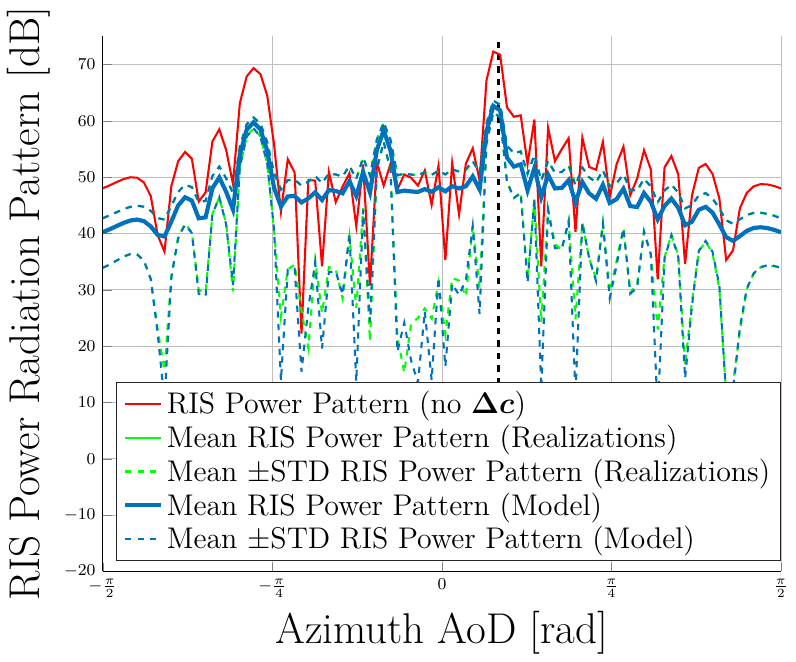}}
    }
    \subfloat[$N=64 \times 64$, AoD: $(15^\circ, 0^\circ)$\label{fig:Fig_RIS_patterns15_N3_plin}]{%
    \resizebox{0.317\textwidth}{!}
      %{\input{Model_validation_high_var/model_val_case15_N3}}
      {\includegraphics[width=\textwidth]{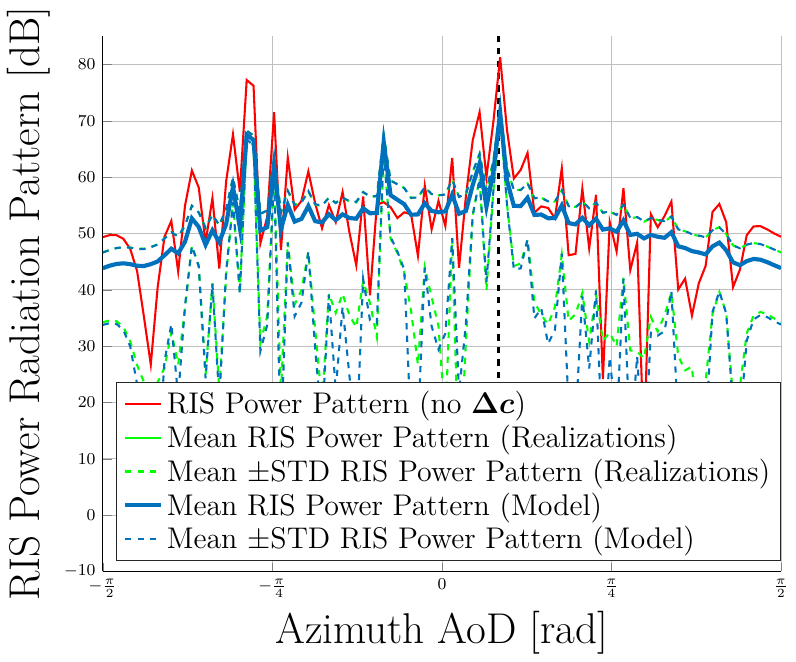}}
    }
\caption{Validation of capacitance variations impact on the RIS radiation pattern under the piecewise-linear model (high variability) assumption. This case supposes that
both realizations and the model (Section~\ref{subsec:Exp_value_high_variations}) assume nominal amplitude and piecewise-linear phase for the RIS element reflection coefficient. The RIS is illuminated from $(-15^\circ, 0^\circ)$ and steered towards $(30^\circ, 0^\circ)$ (top row) and $(15^\circ, 0^\circ)$ (bottom row), with the target AoD indicated by the gray dashed line. The computed RIS configuration is 1-bit quantized and optimized via~\eqref{eq:MRT_solution}. Curves represent: the nominal pattern without variations (\textbf{red}); the empirical mean RIS pattern ($\pm$ STD) from $10000$ Monte Carlo runs under i.i.d. Gaussian variations with $\sigma_c{=}C_{\text{off}}/4$ (\textbf{green}); and the theoretical mean RIS pattern ($\pm$ STD) as per Section~\ref{subsec:Exp_value_high_variations} (\textbf{blue}). Results are sampled at 100 AoDs for each of three RIS sizes, $16{\times}16$, $32{\times}32$, $64{\times}64$.}
\label{fig:Fig_RIS_patterns_varying_N_plin_model_valid}
\end{figure*}

As illustrated in Figs.~\ref{fig:Fig_RIS_patterns_varying_N_lin_model_valid} and~\ref{fig:Fig_RIS_patterns_varying_N_plin_model_valid}, a perfect agreement is achieved between the proposed model and the Monte Carlo simulations when the latter are executed under the exact same theoretical assumptions. Specifically, this match is observed when the reflection amplitude is assumed independent of the capacitance variation, while the phase response is modeled as a linear or piecewise linear function. This precise alignment in Figs.~\ref{fig:Fig_RIS_patterns_varying_N_lin_model_valid} and~\ref{fig:Fig_RIS_patterns_varying_N_plin_model_valid} validates the derived mathematical framework.

Furthermore, it is important to evaluate the predictive accuracy of the proposed models, which substitute the actual nonlinear TLM behavior with a variation-free nominal amplitude and linear or piecewise-linear phase approximations, against the actual nonlinear model. As shown in Figs.~\ref{fig:Fig_RIS_patterns_varying_N_lin_model} and~\ref{fig:Fig_RIS_patterns_varying_N_plin_model}, the approximate models accurately capture the essential radiation trends under hardware imperfections, including main-lobe reduction, existing side-lobe elevation and the emergence of new side-lobes at specific angles. This robust predictive capability renders the proposed model highly suitable for estimating the expected RIS pattern of any given configuration. Despite minor numerical deviations, the fundamental characteristics are preserved, rendering the model highly suitable for robust beamforming (identifying optimized RIS configurations) to ensure predictable main-to-sidelobe gain performance. To rigorously quantify this accuracy, a detailed validation under distinct perturbation regimes, specifically differentiating between small- and large-variance i.i.d. Gaussian variations, is presented in the subsequent subsections.

\subsection{Model Validation and Pattern Optimization under small-variance Additive i.i.d. Gaussian Variations}
For the small-variance variations case, various RIS power patterns are presented in Fig.~\ref{fig:Fig_RIS_patterns_varying_N_lin_model}. All patterns (plots) are computed for the same RIS configuration. It is observed that the theoretical calculations for both the mean RIS power pattern and its STD closely coincide with the TLM-based Monte Carlo realizations. Any minor deviations between the introduced theoretical model and  simulation results are primarily attributed to two factors: the approximation of the reflection amplitude, which is assumed to be independent of the variations, and the inherent approximation errors introduced during the linearization of the TLM phase. Despite these slight discrepancies, both the theoretical model and the Monte Carlo simulations accurately predict how the presence of capacitance variations alters the initial (without varaitions) radiation pattern. Specifically, the variations are shown to cause a reduction in the amplitude of the main lobe---although this effect is nearly negligible for low-variance scenarios---and an increase in the side-lobe levels. In addition, variations appear to cause a subtle shift in some sidelobes and in few cases the generation of new ones even for the low-variance variations case. As a result, the beamforming efficiency is degraded and this degradation needs to be quantified and incorporated into the RIS optimization. 

\begin{figure*}[t!] 

    \centering
    
    \subfloat[$N=32 \times 32$, Mean Patterns\label{fig:Fig_RIS_patterns_B_N1_lin_Greedy}]{%
    \resizebox{0.24\textwidth}{!}
      %{\input{Tikz_Plots_Lvar_noAmpl_Greedy/Fig_RIS_Patterns_N1_linMean}} 
      {\includegraphics[width=\textwidth]{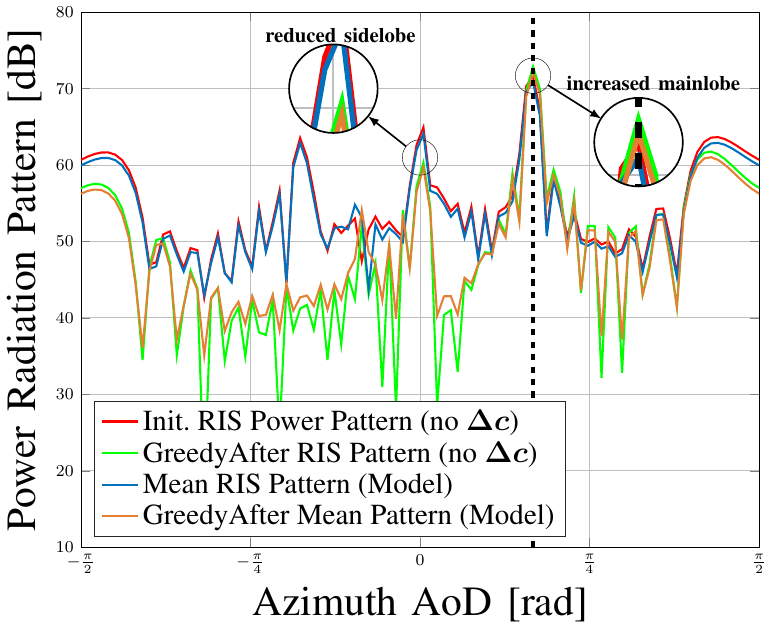}}
    } 
    \subfloat[$N=32 \times 32$, STD Patterns\label{fig:Fig_RIS_patterns_B_N2_lin_Greedy}]{%
    \resizebox{0.24\textwidth}{!}
      %{\input{Tikz_Plots_Lvar_noAmpl_Greedy/Fig_RIS_Patterns_N1_linSTD}}
      {\includegraphics[width=\textwidth]{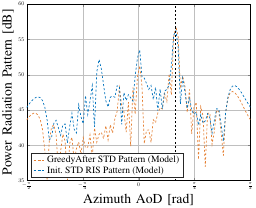}}
    } 
    \subfloat[$N=64 \times 64$, Mean Patterns\label{fig:Fig_RIS_patterns_B_N3_lin_Greedy}]{%
    \resizebox{0.24\textwidth}{!}
      %{\input{Tikz_Plots_Lvar_noAmpl_Greedy/Fig_RIS_Patterns_N2_linMean}}
      {\includegraphics[width=\textwidth]{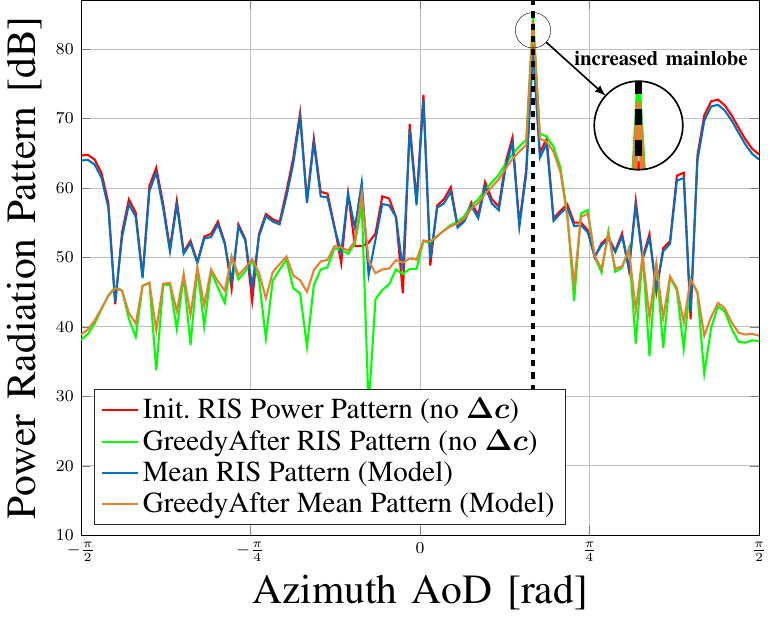}}
    } 
    \subfloat[$N=64 \times64$, STD Patterns\label{fig:Fig_RIS_patterns_B_N4_lin_Greedy}]{%
    \resizebox{0.24\textwidth}{!}
      %{\input{Tikz_Plots_Lvar_noAmpl_Greedy/Fig_RIS_Patterns_N2_linSTD}}
      {\includegraphics[width=\textwidth]{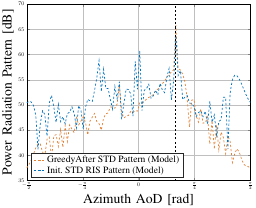}}
    }
    \\
    \subfloat[$N=32 \times 32$, Mean patterns\label{fig:Fig_RIS_patterns_B_N1_plin_Greedy}]{%
    \resizebox{0.24\textwidth}{!}
      %{\input{Tikz_Plots_Hvar_noAmpl_Greedy/Fig_RIS_Patterns_N1_plinMean}} 
      {\includegraphics[width=\textwidth]{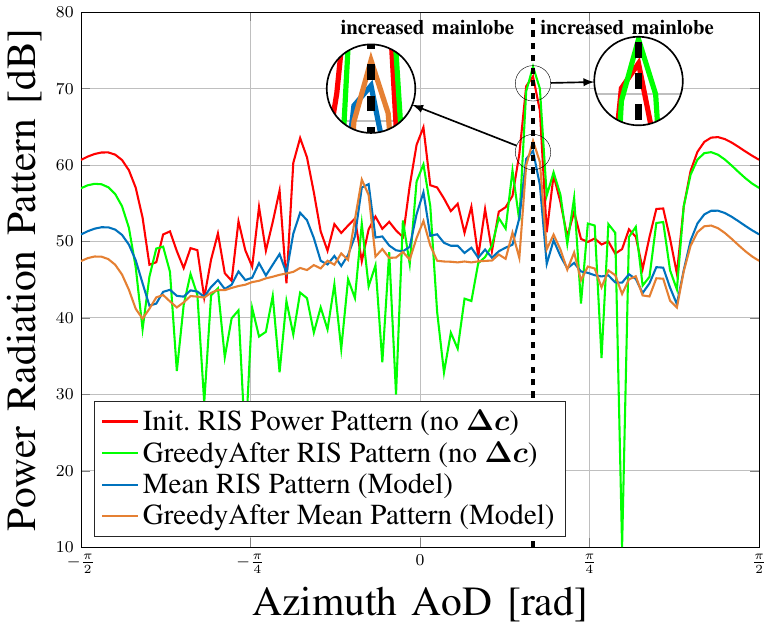}}
    } 
    \subfloat[$N=32 \times 32$, STD patterns\label{fig:Fig_RIS_patterns_B_N2_plin_Greedy}]{%
    \resizebox{0.24\textwidth}{!}
      %{\input{Tikz_Plots_Hvar_noAmpl_Greedy/Fig_RIS_Patterns_N1_plinSTD}}
      {\includegraphics[width=\textwidth]{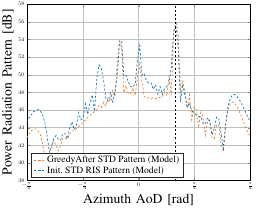}}
    }
    \subfloat[$N=64 \times 64$, Mean patterns\label{fig:Fig_RIS_patterns_B_N3_plin_Greedy}]{%
    \resizebox{0.24\textwidth}{!}
      %{\input{Tikz_Plots_Hvar_noAmpl_Greedy/Fig_RIS_Patterns_N2_plinMean}}
      {\includegraphics[width=\textwidth]{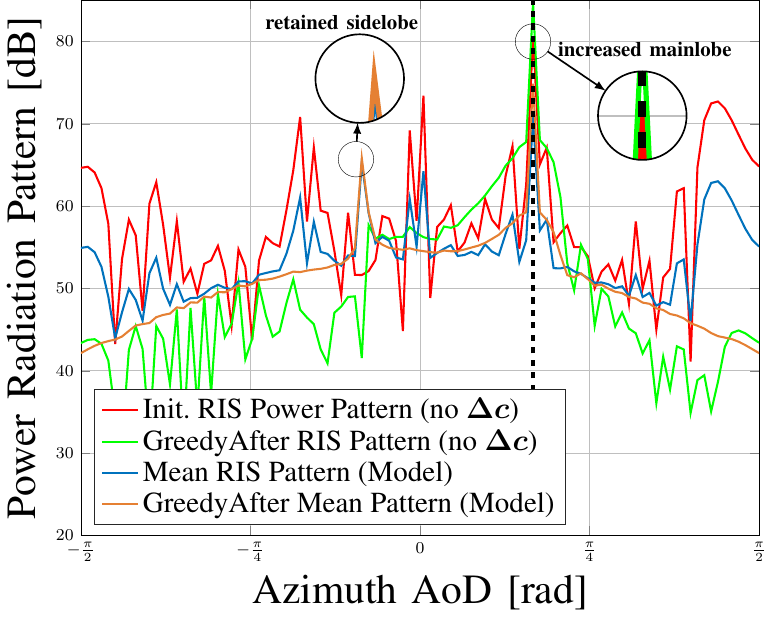}}
    } 
    \subfloat[$N=64 \times64$, STD patterns\label{fig:Fig_RIS_patterns_B_N4_plin_Greedy}]{%
    \resizebox{0.24\textwidth}{!}
      %{\input{Tikz_Plots_Hvar_noAmpl_Greedy/Fig_RIS_Patterns_N2_plinSTD}}
      {\includegraphics[width=\textwidth]{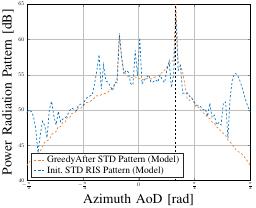}}
    }
    \caption{RIS radiation patterns for a target AoD of $(30^\circ, 0^\circ)$, comparing baseline and greedy-optimized RIS configurations. For both RIS configurations, the results display the mean pattern, the standard deviation pattern, and the RIS pattern under variance-free model. For the RIS configuration computed by~\eqref{eq:MRT_solution} and using 1-bit quantization, \textbf{red curve} depicts the nominal RIS pattern without variations, while \textbf{blue curve} corresponds to the theoretical calculation of the mean/STD RIS power pattern (as per Section~\ref{sec:model_formulation}). \textbf{Orange curve} illustrates the mean RIS pattern optimized using the proposed model in Section~\ref{sec:model_formulation}, via a greedy algorithm with uniform initialization, and \textbf{green curve}, depicts the RIS pattern derived from the variance-free model, but for the occurring after the optimization RIS configuration, to highlight that the proposed model could also be used as an optimization method for the existing (variations-free) model as well. All patterns are sampled at $100$ distinct AoD. Top row Figs., correspond to the linear model of TLM, concerning low per-element capacitance variability, while bottom row Figs. correspond to the piecewise linear model of TLM, concerning high per-element capacitance variability.}
\label{fig:Fig_RIS_patterns_varying_N_lin_Greedy_all}
\end{figure*}

Thus, by accurately modeling the behavior of the RIS under hardware variations, the theoretical model can now be utilized during the optimization process to provide more robust configurations. This approach inherently leads to robust beamforming designs as shown in Fig.~\ref{fig:Fig_RIS_patterns_varying_N_lin_Greedy_all}. Specifically, as shown in Figs.~\ref{fig:Fig_RIS_patterns_B_N1_lin_Greedy}--\ref{fig:Fig_RIS_patterns_B_N4_lin_Greedy}, variation-aware optimization significantly improves the radiation pattern even under low-variance conditions. Specifically, the main-lobe power increases by up to ${\approx} 4\,\text{dB}$ for a $64{\times}64$ RIS configuration. Concurrently, side-lobe levels drop substantially; for large RIS sizes, most side lobes are suppressed by $10$--$20\,\text{dB}$, while the most persistent ones are attenuated by $2$--$3\,\text{dB}$. Remarkably, the greedy algorithm yields superior configurations even when evaluated under variation-free conditions, as the introduced perturbations $\Delta C$ help alleviate the non-convexity of the optimization landscape. Thus, this framework serves as an advanced optimization tool even for perfectly robust, variation-free RIS hardware.

\subsection{Model Validation and Pattern Optimization under large-variance Additive i.i.d. Gaussian Variations}
The large-variance case follows the same radiation trends as the small-variance regime: main-lobe reduction, side-lobe elevation, and emergence of new side-lobes (Fig.~\ref{fig:Fig_RIS_patterns_varying_N_plin_model}). Although degradation is more severe under these harsher conditions, the theoretical model remains valid and closely matches Monte Carlo realizations. Nevertheless, integrating this statistical model into the greedy Algorithm~\ref{alg:greedy_opt} still yields clear performance enhancements (Figs.~\ref{fig:Fig_RIS_patterns_B_N1_plin_Greedy}--\ref{fig:Fig_RIS_patterns_B_N4_plin_Greedy}). Interestingly, after using Algorithm~\ref{alg:greedy_opt}, the newly emerged side lobes remain relatively unaffected, neither increasing nor decreasing significantly. Instead, the achieved main-to-sidelobe gain of ${\approx} 1$--$2\,\text{dB}$ (depending on the RIS size) is driven by a slight main-lobe enhancement combined with the effective suppression of pre-existing side lobes. Furthermore, the optimized configurations exhibit a noticeably reduced STD (Figs.~\ref{fig:Fig_RIS_patterns_B_N2_plin_Greedy} and~\ref{fig:Fig_RIS_patterns_B_N4_plin_Greedy}).

\section{Conclusions}
\label{sec:conclusions}
In this paper, a comprehensive analytical framework for assessing the impact of RIS hardware imperfections on its achievable beamforming capability was presented. In particular, a novel statistical model quantifying the effect of the varactor capacitance fluctuations on the RIS reflection coefficients and the resulting mean RIS power radiation pattern was introduced. The analysis explicitly accounted for independent element-level perturbations, which were modeled as Gaussian random variables. To deal with such inherently introduced imperfections, a robust RIS radiation pattern optimization method, based on a greedy approach, was designed to optimize the derived expression for the mean RIS power radiation pattern. It was demonstrated that, by applying the proposed optimization, the system's robustness against hardware variations is significantly enhanced. Quantitative results indicated that a performance gain of $1$--$4$~dB in the main lobe, coupled with a significant reduction in sidelobe levels, can be achieved. Furthermore, it was demonstrated that the proposed methodology is highly versatile. While the analysis was primarily demonstrated using the TLM, it can be seamlessly adapted to operate with other RIS element models, provided that an appropriate linearization of the reflection coefficient is applied.
  
All in all, the presented framework provides a powerful tool for RIS system designers, offering the capability to analyze and determine which specific RIS size best serves the targeted beamforming gain objectives under realistic hardware imperfections and constraints. Ultimately, this approach effectively replaces time-consuming Monte Carlo simulations, facilitating an efficient and robust design for RIS-assisted communication systems, contributing toward the technology's adoption in future radio-access network infrastructure.
\begin{figure*}[!t]
\normalsize
\begin{equation}
\label{eq:large_eq_cross_plus}
\begin{aligned}
G_{+}(C_i)& {=}  \mathbb{E}_{\Delta C_i} \!\big[\!e^{\jmath \theta(C_i{+}\Delta C_i)}\!\big]=\frac{1}{2} e^{\jmath (b_{22} + a_{22} C_i) - \frac{a_{22}^2 \sigma_c^2}{2}} \left[ \operatorname{erf}\left( \frac{-C_1 + C_i + \jmath a_{22} \sigma_c^2}{\sqrt{2} \sigma_c} \right) - \operatorname{erf}\left( \frac{-C_2 + C_i + \jmath a_{22} \sigma_c^2}{\sqrt{2} \sigma_c} \right) \right] \\
&+ \frac{1}{2} e^{\jmath (b_{23} + a_{23} C_i) - \frac{a_{23}^2 \sigma_c^2}{2}} \left[ 1 {+} \operatorname{erf}\left( \frac{{-}C_2 {+} C_i {+} \jmath a_{23} \sigma_c^2}{\sqrt{2} \sigma_c} \right) \right] 
+ \frac{1}{2} e^{\jmath (b_{21} + a_{21} C_i) - \frac{a_{21}^2 \sigma_c^2}{2}} \operatorname{erfc}\left( \frac{{-}C_1 {+} C_i {+} \jmath a_{21} \sigma_c^2}{\sqrt{2} \sigma_c} \right)
\end{aligned}
\end{equation}
%\noindent \textit{Note:} $\operatorname{erf}(\cdot)$ and $\operatorname{erfc}(\cdot)$ denote the error function and the complementary error function, respectively.
\vspace{-0.43cm}

\hrulefill
\vspace*{4pt}
\end{figure*}
\begin{figure*}[!t]
\vspace{-0.7cm}
\normalsize
\begin{equation}
\label{eq:large_eq_cross_minus}
\begin{aligned}
G_{-}(C_j)&{=} \mathbb{E}_{\Delta C_j}\!\big[\!e^{{-} \jmath \theta(C_j{+}\Delta C_j)}\!\big] =  \frac{1}{2} e^{-\jmath (b_{21} + a_{21} C_j) - \frac{a_{21}^2 \sigma_{c}^2}{2}} \left[ 1 + \operatorname{erf}\left( \frac{C_1 - C_j + \jmath a_{21} \sigma_{c}^2}{\sqrt{2} \sigma_{c}} \right) \right] {+} \frac{1}{2} e^{-\jmath (b_{22} + a_{22} C_j) - \frac{a_{22}^2 \sigma_{c}^2}{2}} \\ &\left[ \operatorname{erf}\left( \frac{C_2 {-} C_j {+} \jmath a_{22} \sigma_{c}^2}{\sqrt{2} \sigma_{c}} \right) {-} \operatorname{erf}\left( \frac{C_1 {-} C_j {+} j a_{22} \sigma_{c}^2}{\sqrt{2} \sigma_{c}} \right) \right] {+} \frac{1}{2} e^{-\jmath (b_{23} + a_{23} C_j) - \frac{a_{23}^2 \sigma_{c}^2}{2}} \operatorname{erfc}\left( \frac{C_2 {-} C_j + \jmath a_{23} \sigma_{c}^2}{\sqrt{2} \sigma_{c}} \right)
\end{aligned}
\end{equation}
%\noindent \textit{Note:} $\operatorname{erf}(\cdot)$ and $\operatorname{erfc}(\cdot)$ denote the error function and the complementary error function, respectively.

\vspace{-0.1cm}
\hrulefill
\vspace*{4pt}
\end{figure*}
\appendices
\section{Integrals Required for Section~\ref{subsec:Exp_value_high_variations}}\label{appendix:2}
Integrals $G_{+}(C_i)$ and $G_{-}(C_j)$  are shown in~\eqref{eq:large_eq_cross_plus} and~\eqref{eq:large_eq_cross_minus}.

\section{Gaussian Random Variable Property}
\label{appendix:1}
Let $X$ be a random variable that follows a zero-mean Gaussian PDF, $
f_{X}(x) = \big(\sqrt{2\pi\sigma^2}\big)^{-1} \, e^{%\Big(
-\frac{x^2}{2\sigma^2}},$
where $\sigma^2$ denotes the variance of the PDF. Then, for any arbitrary constant $k$, the following equality holds
\begin{align}
    &\mathbb{E}_{X}\!\left[e^{\jmath kX}\right]
= e^{-\frac{1}{2}k^2\sigma^2} \label{eq:property_eqs1}%\\
%& \mathbb{E}_{X}\!\left[X e^{\jmath kX}\right] = e^{-\frac{1}{2}k^2\sigma^2} \jmath k\sigma^2
%\label{eq:property_eqs2}
\end{align}
\begin{proof}
Eq.~\eqref{eq:property_eqs1} holds since
\begin{align}
    &\mathbb{E}_{X}\!\left[e^{\jmath kX}\right]
= \int_{-\infty}^{\infty} e^{\jmath kx}\;
\frac{1}{\sqrt{2\pi\sigma^2}}e^{-\frac{x^2}{2\sigma^2}}\,dx =\\
&= \int_{-\infty}^{\infty}
\frac{1}{\sqrt{2\pi\sigma^2}}e^{-\frac{1}{2\sigma^2} (x^2 -\jmath 2 \sigma^2 kx)}\,dx \\
&= \int_{-\infty}^{\infty}
\frac{1}{\sqrt{2\pi\sigma^2}}e^{-\frac{1}{2\sigma^2} \big(x^2 -2x\jmath k \sigma^2 + (\jmath k\sigma^2)^2 - (\jmath k\sigma^2)^2 \big)}\,dx \\
&= e^{\frac{(\jmath k\sigma^2)^2}{2\sigma^2}}\int_{-\infty}^{\infty}
\frac{1}{\sqrt{2\pi\sigma^2}}e^{-\frac{(x-\jmath k\sigma^2)^2 }{2\sigma^2} }\,dx = e^{-\frac{1}{2}k^2\sigma^2}.
\end{align}
This completes the proof.
\end{proof}

\IEEEtriggercmd{\enlargethispage{-1ex}}
\IEEEtriggeratref{10}
\bibliographystyle{IEEEtran}
\bibliography{ris_refs}

@ARTICLE{huang2019reconfigurable,
  author={Huang, Chongwen and Zappone, Alessio and Alexandropoulos, George C. and Debbah, Mérouane and Yuen, Chau},
  journal={IEEE Trans. Wireless Commun.}, 
  title={{R}econfigurable Intelligent Surfaces for Energy Efficiency in Wireless Communication},
  month = aug,
  year={2019},
  volume={18},
  number={8},
  pages={4157-4170}
}

@article{strinati2021wireless,
  title={{R}econfigurable, intelligent, and sustainable wireless environments for 6{G} smart connectivity},
  author={Calvanese Strinati, Emilio and Alexandropoulos, George C and Wymeersch, Henk and Denis, Benoit and Sciancalepore, Vincenzo and d'Errico, Raffaele and Clemente, Antonio and Phan-Huy, Dinh-Thuy and De Carvalho, Elisabeth and Popovski, Petar},
  journal={IEEE Commun. Mag.},
  volume={59},
  number={10},
  pages={99--105},
  year={2021},
  publisher={IEEE}
}

@ARTICLE{10243495,
  author={Chepuri, Sundeep Prabhakar and Shlezinger, Nir and Liu, Fan and Alexandropoulos, George C. and Buzzi, Stefano and Eldar, Yonina C.},
  journal={IEEE Signal Process. Mag.}, 
  title={Integrated Sensing and Communications With Reconfigurable Intelligent Surfaces: From signal modeling to processing}, 
  year={2023},
  volume={40},
  number={6},
  pages={41-62}
  }

@ARTICLE{9847080,
  author={Pan, Cunhua and Zhou, Gui and Zhi, Kangda and Hong, Sheng and Wu, Tuo and Pan, Yijin and Ren, Hong and Renzo, Marco Di and Lee Swindlehurst, A. and Zhang, Rui and Zhang, Angela Yingjun},
  journal={IEEE J. Sel. Topics Signal Process.}, 
  title={{A}n Overview of Signal Processing Techniques for {RIS/IRS}-Aided Wireless Systems}, 
  year={2022},
  volume={16},
  number={5},
  pages={883-917},
  doi={10.1109/JSTSP.2022.3195671}}

@inproceedings{RDK21,
  author={Rahal, Moustafa and Denis, Benoît and Keykhosravi, Kamran and Uguen, Bernard and Wymeersch, Henk},
  booktitle={Proc. IEEE SPAWC}, 
  address={Lucca, Italy},
  title={{RIS}-Enabled Localization Continuity Under Near-Field Conditions}, 
  year={2021},
  pages={436-440}
}

@article{Tsinghua_RIS_Tutorial,
  title={{Reconfigurable intelligent surfaces for wireless communications: Overview of hardware designs, channel models, and estimation techniques}},
  author={Jian, Mengnan and Alexandropoulos, George C and Basar, Ertugrul and Huang, Chongwen and Liu, Ruiqi and Liu, Yuanwei and Yuen, Chau},
  journal={Intell. Converged Netw.},
  volume={3},
  number={1},
  pages={1--32},
  year={2022},
  publisher={TUP}
}

@article{cui2019secure,
  title={{Secure wireless communication via intelligent reflecting surface}},
  author={Cui, Miao and Zhang, Guangchi and Zhang, Rui},
  journal={IEEE Wireless Commun. Lett.},
  volume={8},
  number={5},
  pages={1410--1414},
  year={2019},
  publisher={IEEE}
}

@article{guan2020intelligent,
  title={{Intelligent reflecting surface assisted secrecy communication: Is artificial noise helpful or not?}},
  author={Guan, Xinrong and Wu, Qingqing and Zhang, Rui},
  journal={IEEE Wireless Commun. Lett.},
  volume={9},
  number={6},
  pages={778--782},
  year={2020},
  publisher={IEEE}
}

@inproceedings{vordonis2025evaluating,
  title={Evaluating Beam Sweeping for {A}o{A} Estimation with an {RIS} Prototype: {I}ndoor/Outdoor Field Trials},
  author={Vordonis, Dimitris and Kompostiotis, Dimitris and Paliouras, Vassilis and Alexandropoulos, George C and Grec, Florin},
  booktitle={Proc. IEEE WCNC},
  pages={1--6},
  year={2025},
  address={Milan, Italy}
}

@inproceedings{kompostiotis2025optimizing,
  title={{O}ptimizing Indoor {RIS}-Aided Physical Layer Security: {A} Codebook-Generation Methodology and Measurement-Based Analysis},
  author={Kompostiotis, Dimitris and Vordonis, Dimitris and Paliouras, Vassilis and Alexandropoulos, George C},
  booktitle={Proc. IEEE PIMRC},
  pages={1--6},
  year={2025},
  address={Istanbul, Turkey}
}

@article{rahal2022arbitrary,
  title={Performance of {RIS}-aided nearfield localization under beams approximation from real hardware characterization},
  author={M. Rahal and B. Denis and K. Keykhosravi and M. F. Keskin and B. Uguen and G. C. Alexandropoulos and H. Wymeersch},
  journal={EURASIP J. Wireless Commun. Netw.},
  volume={86},
  pages={1--23},
  year={2023}
}

@article{cai2020practical,
  title={Practical modeling and beamforming for intelligent reflecting surface aided wideband systems},
  author={Cai, Wenhao and Li, Hongyu and Li, Ming and Liu, Qian},
  journal={IEEE Commun. Lett.},
  volume={24},
  number={7},
  pages={1568--1571},
  year={2020},
  publisher={IEEE}
}

@inproceedings{ramezani2023broad,
  title={Broad beam reflection for {RIS}-assisted {MIMO} systems with planar arrays},
  author={Ramezani, Parisa and Girnyk, Maksym A and Bjornson, Emil},
  booktitle={Proc. Asilomar Conf. Signals, Sys., Comp.},
  pages={504--508},
  year={2023},
  address={Pacific Grove, USA}
}

@inproceedings{MIMO_1bit,
  title={{MIMO} communications with 1-bit {RIS}: Asymptotic analysis and over-the-air channel diagonalization},
  author={P. Gavriilidis and K. Stylianopoulos and and G. C. Alexandropoulos},
  booktitle={Proc. Asilomar Conf. Signals, Sys., Comp.},
  pages={1467--1474},
  year={2025},
  address={Pacific Grove, USA}
}

@article{ramezani2023dual,
  title={Dual-polarized reconfigurable intelligent surface-assisted broad beamforming},
  author={Ramezani, Parisa and Girnyk, Maksym A and Bj{\"o}rnson, Emil},
  journal={IEEE Commun. Lett.},
  volume={27},
  number={11},
  pages={3073--3077},
  year={2023},
  publisher={IEEE}
}

@article{liu2023near,
  title={{Near-field communications: A tutorial review}},
  author={Liu, Yuanwei and Wang, Zhaolin and Xu, Jiaqi and Ouyang, Chongjun and Mu, Xidong and Schober, Robert},
  journal={IEEE Open J. Commun. Society},
  volume={4},
  pages={1999--2049},
  year={2023},
  publisher={IEEE}
}

@ARTICLE{11030582,
  author={Arslan, Mehmet Emin and Nordmeyer, Ulrich and Neumann, Niels},
  journal={IEEE Microwave Wireless Technol. Lett.}, 
  title={General Method for Characterizing Switchable Elements for RIS Using De-Embedding Structures}, 
  year={2025},
  volume={35},
  number={9},
  pages={1444-1447},
  doi={10.1109/LMWT.2025.3574501}}

@misc{Rains_OpenRIS_Github,
  author = {Rains, James and others},
  title = {{OpenRIS}: {A}n Open-Source Reconfigurable Intelligent Surface},
  year = {2022},
  publisher = {GitHub},
  journal = {GitHub repository},
  howpublished = {\url{https://github.com/jimrains/OpenRIS}}}

@article{akaike2004analysis,
  title={An analysis of nonlinear terms in capacitance-voltage characteristic for anti-series-connected varactor-diode pair},
  author={Akaike, Masami and Ohira, Takashi and Inagaki, Keizo and Han, Qing},
  journal={Int. J. RF Microwave Comp.-Aided Engineering},
  volume={14},
  number={3},
  pages={274--282},
  year={2004},
  publisher={Wiley Online Library}
}

@article{buisman2012rf,
  title={{RF} power insensitive varactors},
  author={Buisman, Koen and Huang, Cong and Zampardi, Peter J and de Vreede, Leo CN},
  journal={IEEE Microwave Wireless Compon. Lett.},
  volume={22},
  number={8},
  pages={418--420},
  year={2012},
  publisher={IEEE}
}

@manual{Skyworks_SMV1408_Datasheet,
  organization = {Skyworks Solutions, Inc.},
  title        = {{SMV1405} to {SMV1430} series: {P}lastic-packaged abrupt junction tuning varactors},
  year         = {2020},
  month        = {January},
  number       = {200068W},
  note         = {Data Sheet}
}

@article{li2003robust,
  title={{On robust Capon beamforming and diagonal loading}},
  author={Li, Jian and Stoica, Petre and Wang, Zhisong},
  journal={IEEE Trans. Signal Process.},
  volume={51},
  number={7},
  pages={1702--1715},
  year={2003},
  publisher={IEEE}
}

@article{vorobyov2003robust,
  title={{Robust adaptive beamforming using worst-case performance optimization: A solution to the signal mismatch problem}},
  author={Vorobyov, Sergiy A and Gershman, Alex B and Luo, Zhi-Quan},
  journal={IEEE Trans. Signal Process.},
  volume={51},
  number={2},
  pages={313--324},
  year={2003},
  publisher={IEEE}
}

@article{lorenz2005robust,
  title={Robust minimum variance beamforming},
  author={Lorenz, Robert G and Boyd, Stephen P},
  journal={IEEE Trans. Signal Process.},
  volume={53},
  number={5},
  pages={1684--1696},
  year={2005},
  publisher={IEEE}
}

@article{wu2021intelligent,
  title={Intelligent reflecting surface-aided wireless communications: {A} tutorial},
  author={Wu, Qingqing and Zhang, Shuowen and Zheng, Beixiong and You, Changsheng and Zhang, Rui},
  journal={IEEE Trans. Commun.},
  volume={69},
  number={5},
  pages={3313--3351},
  year={2021},
  publisher={IEEE}
}

@inproceedings{RIS_1bit,
  title={Asymptotically optimal closed-form phase configuration of 1-bit {RISs} via sign alignment},
  author={K. Stylianopoulos and P. Gavriilidis and G. C. Alexandropoulos},
  booktitle={Proc. IEEE SPAWC},
  address={Lucca, Italy},
  year={2024}
}

@article{costa2021electromagnetic,
  title={Electromagnetic model of reflective intelligent surfaces},
  author={Costa, Filippo and Borgese, Michele},
  journal={IEEE Open J. Commun. Society},
  volume={2},
  pages={1577--1589},
  year={2021},
  publisher={IEEE}
}

@article{abeywickrama2020intelligent,
  title={Intelligent reflecting surface: Practical phase shift model and beamforming optimization},
  author={Abeywickrama, Samith and Zhang, Rui and Wu, Qingqing and Yuen, Chau},
  journal={IEEE Trans. Commun.},
  volume={68},
  number={9},
  pages={5849--5863},
  year={2020},
  publisher={IEEE}
}

@article{bjornson2022reconfigurable,
  title={{Reconfigurable intelligent surfaces: A signal processing perspective with wireless applications}},
  author={Bj{\"o}rnson, Emil and Wymeersch, Henk and Matthiesen, Bho and Popovski, Petar and Sanguinetti, Luca and de Carvalho, Elisabeth},
  journal={IEEE Signal Process. Mag.},
  volume={39},
  number={2},
  pages={135--158},
  year={2022},
  publisher={IEEE}
}

@inproceedings{bjornson2021optimizing,
  title={Optimizing a binary intelligent reflecting surface for {OFDM} communications under mutual coupling},
  author={Bj{\"o}rnson, Emil},
  booktitle={Proc. Int. Workshop Smart Ant.}, 
  pages={1--6},
  year={2021},
  organization={VDE}
}

@inproceedings{kompostiotis2023secrecy,
  title={{S}ecrecy Rate Maximization in {RIS}-Enabled {OFDM} Wireless Communications: {T}he Circuit-Based Reflection Model Case},
  author={Kompostiotis, Dimitris and Vordonis, Dimitris and Paliouras, Vassilis and Alexandropoulos, George C},
  booktitle={Proc. IEEE ICC Workshops},
  pages={1529--1534},
  year={2023},
  address={Rome, Italy}
}

@article{pan2020multicell,
  title={Multicell {MIMO} communications relying on intelligent reflecting surfaces},
  author={Pan, Cunhua and Ren, Hong and Wang, Kezhi and Xu, Wei and Elkashlan, Maged and Nallanathan, Arumugam and Hanzo, Lajos},
  journal={IEEE Trans. Wireless Commun.},
  volume={19},
  number={8},
  pages={5218--5233},
  year={2020},
  publisher={IEEE}
}

@inproceedings{kompostiotis2024evaluation,
  title={{E}valuation of {RIS}-enabled {B5G/6G} indoor positioning and mapping using ray tracing models},
  author={Kompostiotis, Dimitris and Vordonis, Dimitris and Paliouras, Vassilis and Alexandropoulos, George C and Grec, Florin},
  booktitle={Proc. Workshop Satellite Navigation Technol.},
  address={Noordwijk, Netherlands},
  year={2024}
}

@inproceedings{kompostiotis2023received,
  title={{Received power maximization with practical phase-dependent amplitude response in RIS-aided OFDM wireless communications}},
  author={Kompostiotis, Dimitris and Vordonis, Dimitris and Paliouras, Vassilis},
  booktitle={Proc. IEEE ICASSP 2023},
  address={Rhodes, Greece},
  year={2023}
}

@ARTICLE{RIS_loc,
	author={K. Keykhosravi and B. Denis and G. C. Alexandropoulos and Z. S. He and A. Albanese and V. Sciancalepore and H. Wymeersch},
	journal={IEEE Veh. Technol. Mag.}, 
	title={Leveraging {RIS}-enabled smart signal propagation for solving infeasible localization problems}, 
	month = {Jun.},
	year={2023},
	volume={18},
	number={2},
	pages={20--28}
}

@ARTICLE{basar2023RIS_mag,
  author={Basar, Ertugrul and Alexandropoulos, George C. and Liu, Yuanwei and Wu, Qingqing and Jin, Shi and Yuen, Chau and Dobre, Octavia A. and Schober, Robert},
  journal={IEEE Veh. Technol. Mag.}, 
  title={{R}econfigurable Intelligent Surfaces for 6{G}: {E}merging Hardware Architectures, Applications, and Open Challenges}, 
  year={2024},
  volume={19},
  number={3},
  pages={27-47},
  doi={10.1109/MVT.2024.3415570}}

@article{alexandropoulos2023ris,
  title={{RIS}-enabled smart wireless environments: {D}eployment scenarios, network architecture, bandwidth and area of influence},
  author={Alexandropoulos, George C and others},
  journal={EURASIP J. Wireless Commun. Netw.},
  volume={2023},
  number={1},
  pages={103},
  year={2023},
  publisher={Springer}
}

@ARTICLE{RIS_smart_cities,
  author={H. Chen and H. Kim and M. Ammous and G. Seco-Granados and G. C. Alexandropoulos and S. Valaee and H. Wymeersch},
  journal={IEEE Commun. Mag.}, 
  title={{RISs} and sidelink communications in smart cities: The key to seamless localization and sensing}, 
  month={Aug.},
  year={2023},
  volume={61},
  number={8},
  pages={140-146}
}

@article{wymeersch2022localisation,
  title={Localisation and sensing use cases and gap analysis},
  author={Wymeersch, Henk and others},
  journal={Hexa-X project Deliverable D},
  volume={3},
  year={2022},
  url={https://hexa-x.eu/wp-content/uploads/2022/01/Hexa-X-D3.1_v1.4.pdf}
}

@ARTICLE{10005197,
  author={Zarini, Hosein and Gholipoor, Narges and Mili, Mohammad Robat and Rasti, Mehdi and Tabassum, Hina and Hossain, Ekram},
  journal={IEEE Trans. Commun.}, 
  title={{R}esource Management for Multiplexing e{MBB} and {URLLC} Services Over {RIS}-Aided {TH}z Communication}, 
  year={2023},
  volume={71},
  number={2},
  pages={1207-1225},
  doi={10.1109/TCOMM.2023.3233988}}

@ARTICLE{10989512,
  author={Li, Yongxiao and Khan, Feroz and Ahmed, Manzoor and Soofi, Aized Amin and Khan, Wali Ullah and Sheemar, Chandan Kumar and Asif, Muhammad and Han, Zhu},
  journal={IEEE Internet of Things J.}, 
  title={{RIS}-Based Physical Layer Security for Integrated Sensing and Communication: {A} Comprehensive Survey}, 
  year={2025},
  volume={12},
  number={16},
  pages={32444-32468},
  doi={10.1109/JIOT.2025.3567553}}

@ARTICLE{9852716,
  author={Luo, Honghao and Liu, Rang and Li, Ming and Liu, Yang and Liu, Qian},
  journal={IEEE Trans. Veh. Technol.}, 
  title={{J}oint Beamforming Design for {RIS}-Assisted Integrated Sensing and Communication Systems}, 
  year={2022},
  volume={71},
  number={12},
  pages={13393-13397},
  doi={10.1109/TVT.2022.3197448}}

@article{wang2025impact,
  title={On the Impact of Phase Errors in Phase-Dependent Amplitudes of Near-Field {RIS}s},
  author={Wang, Ke and Lam, Chan-Tong and Ng, Benjamin K and Liu, Yue},
  journal={IEEE Trans. Veh. Technol.},
  volume={75},
  number={6},
  pages={10826--10842},
  year={2025},
}

@INPROCEEDINGS{765552,
  author={Lo, T.K.Y.},
  booktitle={1999 IEEE International Conference on Communications (Cat. No. 99CH36311)}, 
  title={Maximum ratio transmission}, 
  year={1999},
  volume={2},
  number={},
  pages={1310-1314 vol.2},
  doi={10.1109/ICC.1999.765552}}

@misc{skywater_pdk,
  author       = {{SkyWater Technology and Google}},
  title        = {{SkyWater SKY130 process design kit (PDK)}},
  year         = {2020},
  publisher    = {GitHub},
  journal      = {GitHub Repository},
  howpublished = {\url{https://github.com/google/skywater-pdk}},
  note         = {Accessed: June 2026}
}

\end{document}